\documentclass[a4paper]{jpconf}
\usepackage{graphicx}
\usepackage{ifthen} 
\usepackage{lineno}
\usepackage{xspace}
\usepackage{amsmath} 
\usepackage{amssymb}
\usepackage{amsfonts}
\usepackage{upgreek} 

\usepackage{hyperxmp}
\usepackage{hyperref}

\usepackage{cite} 
\usepackage{mciteplus}

\newboolean{pdflatex}
\setboolean{pdflatex}{true} 

\newboolean{articletitles}
\setboolean{articletitles}{true} 
\newboolean{uprightparticles}
\setboolean{uprightparticles}{false} 
\newboolean{inbibliography}
\setboolean{inbibliography}{false} 

\usepackage{xspace} 
\usepackage{upgreek}

\def\lhcb   {\mbox{LHCb}\xspace}

\def\MagUp {\mbox{\em Mag\kern -0.05em Up}\xspace}

\ifthenelse{\boolean{uprightparticles}}%
{

 \def\Peta        {\ensuremath{\upeta}\xspace}

 \def\Pmu         {\ensuremath{\upmu}\xspace}

 \def\Ppi         {\ensuremath{\uppi}\xspace}

 \def\Pchi        {\ensuremath{\upchi}\xspace}                 
 \def\Ppsi        {\ensuremath{\uppsi}\xspace}

 \def\PDelta      {\ensuremath{\Delta}\xspace}                 
 \def\PXi         {\ensuremath{\Xi}\xspace}                 
 \def\PLambda     {\ensuremath{\Lambda}\xspace}                 
 \def\PSigma      {\ensuremath{\Sigma}\xspace}                 
 \def\POmega      {\ensuremath{\Omega}\xspace}                 
 \def\PUpsilon    {\ensuremath{\Upsilon}\xspace}

 \def\PB      {\ensuremath{\mathrm{B}}\xspace}                 
                  
 \def\PD      {\ensuremath{\mathrm{D}}\xspace}

 \def\PJ      {\ensuremath{\mathrm{J}}\xspace}                 
 \def\PK      {\ensuremath{\mathrm{K}}\xspace}

 \def\Pb      {\ensuremath{\mathrm{b}}\xspace}                 
 \def\Pc      {\ensuremath{\mathrm{c}}\xspace}                 
                  
 \def\Pe      {\ensuremath{\mathrm{e}}\xspace}

 \def\Pi      {\ensuremath{\mathrm{i}}\xspace}

 \def\Pp      {\ensuremath{\mathrm{p}}\xspace}

}
{

 \def\Peta        {\ensuremath{\eta}\xspace}

 \def\Pmu         {\ensuremath{\mu}\xspace}

 \def\Ppi         {\ensuremath{\pi}\xspace}

 \def\Pchi        {\ensuremath{\chi}\xspace}                 
 \def\Ppsi        {\ensuremath{\psi}\xspace}                 
                  
 \mathchardef\PDelta="7101
 \mathchardef\PXi="7104
 \mathchardef\PLambda="7103
 \mathchardef\PSigma="7106
 \mathchardef\POmega="710A
 \mathchardef\PUpsilon="7107
                  
 \def\PB      {\ensuremath{B}\xspace}                 
                  
 \def\PD      {\ensuremath{D}\xspace}

 \def\PJ      {\ensuremath{J}\xspace}                 
 \def\PK      {\ensuremath{K}\xspace}

 \def\Pb      {\ensuremath{b}\xspace}                 
 \def\Pc      {\ensuremath{c}\xspace}                 
                  
 \def\Pe      {\ensuremath{e}\xspace}

 \def\Pi      {\ensuremath{i}\xspace}

 \def\Pp      {\ensuremath{p}\xspace}

}

\makeatletter
\ifcase \@ptsize \relax
  \newcommand{\miniscule}{\@setfontsize\miniscule{4}{5}}
\or
  \newcommand{\miniscule}{\@setfontsize\miniscule{5}{6}}
\or
  \newcommand{\miniscule}{\@setfontsize\miniscule{5}{6}}
\fi
\makeatother

\DeclareRobustCommand{\optbar}[1]{\shortstack{{\miniscule (\rule[.5ex]{1.25em}{.18mm})}
  \\ [-.7ex] $#1$}}

\def\electron   {{\ensuremath{\Pe}}\xspace}
\def\en         {{\ensuremath{\Pe^-}}\xspace}   

\def\mumu       {{\ensuremath{\Pmu^+\Pmu^-}}\xspace}

\def\cquark    {{\ensuremath{\Pc}}\xspace}
\def\cquarkbar {{\ensuremath{\overline \cquark}}\xspace}
\def\ccbar     {{\ensuremath{\cquark\cquarkbar}}\xspace}
\def\bquark    {{\ensuremath{\Pb}}\xspace}

\def\pion   {{\ensuremath{\Ppi}}\xspace}

\def\pip    {{\ensuremath{\pion^+}}\xspace}
\def\pim    {{\ensuremath{\pion^-}}\xspace}

\def\kaon    {{\ensuremath{\PK}}\xspace}
  \def\Kbar    {{\kern 0.2em\overline{\kern -0.2em \PK}{}}\xspace}

\def\KorKbar    {\kern 0.18em\optbar{\kern -0.18em K}{}\xspace}

\def\Kp      {{\ensuremath{\kaon^+}}\xspace}

  \def\Dbar    {{\kern 0.2em\overline{\kern -0.2em \PD}{}}\xspace}
\def\D       {{\ensuremath{\PD}}\xspace}

\def\DorDbar    {\kern 0.18em\optbar{\kern -0.18em D}{}\xspace}
\def\Dz      {{\ensuremath{\D^0}}\xspace}
\def\Dzb     {{\ensuremath{\Dbar{}^0}}\xspace}

\def\Dm      {{\ensuremath{\D^-}}\xspace}

\def\B       {{\ensuremath{\PB}}\xspace}
\def\Bbar    {{\ensuremath{\kern 0.18em\overline{\kern -0.18em \PB}{}}}\xspace}

\def\BorBbar    {\kern 0.18em\optbar{\kern -0.18em B}{}\xspace}
\def\Bz      {{\ensuremath{\B^0}}\xspace}

\def\Bu      {{\ensuremath{\B^+}}\xspace}

\def\Bp      {{\ensuremath{\Bu}}\xspace}

\def\jpsi     {{\ensuremath{{\PJ\mskip -3mu/\mskip -2mu\Ppsi\mskip 2mu}}}\xspace}
\def\psitwos  {{\ensuremath{\Ppsi{(2S)}}}\xspace}

\def\etac     {{\ensuremath{\Peta_\cquark}}\xspace}

\def\Y#1S{\ensuremath{\PUpsilon{(#1S)}}\xspace}

\def\chic  {{\ensuremath{\Pchi_{c}}}\xspace}

\def\proton      {{\ensuremath{\Pp}}\xspace}
\def\antiproton  {{\ensuremath{\overline \proton}}\xspace}

\def\Lz          {{\ensuremath{\PLambda}}\xspace}
\def\Lbar        {{\ensuremath{\kern 0.1em\overline{\kern -0.1em\PLambda}}}\xspace}
\def\LorLbar     {\kern 0.18em\optbar{\kern -0.18em \PLambda}{}\xspace}

\def\Lb      {{\ensuremath{\Lz^0_\bquark}}\xspace}

\def\Lc      {{\ensuremath{\Lz^+_\cquark}}\xspace}

\def\to                 {\ensuremath{\rightarrow}\xspace}

\def\AT#1     {\ensuremath{A_{\mathrm{T}}^{#1}}\xspace}           

\def\C#1      {\ensuremath{\mathcal{C}_{#1}}\xspace}                       
\def\Cp#1     {\ensuremath{\mathcal{C}_{#1}^{'}}\xspace}                    
\def\Ceff#1   {\ensuremath{\mathcal{C}_{#1}^{\mathrm{(eff)}}}\xspace}        
\def\Cpeff#1  {\ensuremath{\mathcal{C}_{#1}^{'\mathrm{(eff)}}}\xspace}       
\def\Ope#1    {\ensuremath{\mathcal{O}_{#1}}\xspace}                       
\def\Opep#1   {\ensuremath{\mathcal{O}_{#1}^{'}}\xspace}                    

\newcommand{\tev}{\ifthenelse{\boolean{inbibliography}}{\ensuremath{~T\kern -0.05em eV}}{\ensuremath{\mathrm{\,Te\kern -0.1em V}}}\xspace}
\newcommand{\gev}{\ensuremath{\mathrm{\,Ge\kern -0.1em V}}\xspace}
\newcommand{\mev}{\ensuremath{\mathrm{\,Me\kern -0.1em V}}\xspace}
\newcommand{\kev}{\ensuremath{\mathrm{\,ke\kern -0.1em V}}\xspace}
\newcommand{\ev}{\ensuremath{\mathrm{\,e\kern -0.1em V}}\xspace}
\newcommand{\gevc}{\ensuremath{{\mathrm{\,Ge\kern -0.1em V\!/}c}}\xspace}
\newcommand{\mevc}{\ensuremath{{\mathrm{\,Me\kern -0.1em V\!/}c}}\xspace}
\newcommand{\gevcc}{\ensuremath{{\mathrm{\,Ge\kern -0.1em V\!/}c^2}}\xspace}
\newcommand{\gevgevcccc}{\ensuremath{{\mathrm{\,Ge\kern -0.1em V^2\!/}c^4}}\xspace}
\newcommand{\mevcc}{\ensuremath{{\mathrm{\,Me\kern -0.1em V\!/}c^2}}\xspace}

\def\invnb {\ensuremath{\mbox{\,nb}^{-1}}\xspace}

\def\ms   {\ensuremath{{\mathrm{ \,ms}}}\xspace}

\def\gsim{{~\raise.15em\hbox{$>$}\kern-.85em
          \lower.35em\hbox{$\sim$}~}\xspace}
\def\lsim{{~\raise.15em\hbox{$<$}\kern-.85em
          \lower.35em\hbox{$\sim$}~}\xspace}

\def\sqsnn {\ensuremath{\protect\sqrt{s_{\scriptscriptstyle\rm NN}}}\xspace}
\def\pt         {\ensuremath{p_{\mathrm{T}}}\xspace}
\def\ptot       {\ensuremath{p}\xspace}

\def\tell1  {TELL1\xspace}
\def\ukl1   {UKL1\xspace}

\def\sqsnn   {\ensuremath{\protect\sqrt{s_{\scriptscriptstyle\rm NN}}}\xspace}
\def\pA {\proton-A\xspace}
\def\pHe{\proton-He\xspace}
\def\pNe{\proton-Ne\xspace}
\def\pAr{\proton-Ar\xspace}
\def\pPb{\proton-Pb\xspace}
\def\Pbp{Pb-\proton\xspace}
\def\PbPb{Pb-Pb\xspace}
\def\pp{\proton-\proton}
\def\pbar{\antiproton}

\def\pe{\proton-\en\xspace}

\begin{document}
\title{Results on heavy ion physics at \lhcb}
\author{Giacomo Graziani, on behalf of the \lhcb Collaboration}
\address{INFN, Sezione di Firenze}
\ead{Giacomo.Graziani@fi.infn.it}
\begin{abstract}
In the last years, the \lhcb experiment established itself as an
important contributor to heavy ion physics by exploiting some of 
its specific features. Production of particles, 
notably heavy flavour states, can
be studied in \pp, \pPb and Pb-Pb collisions  at LHC energies in the
forward rapidity region (pseudorapidity between 2 and 5), providing
measurements which are highly complementary to the other LHC 
experiments. Moreover, owing to its forward geometry, the detector is 
also well suited to study fixed-target collisions, obtained by impinging
the LHC beams on gas targets with different mass numbers. In this
configuration, \pA collisions can be studied at the relatively 
unexplored scale of \sqsnn$\sim$ 100~GeV, also providing valuable 
inputs to cosmic ray physics.
An overview of the measurements obtained so far by the \lhcb ion program is presented.
\end{abstract}

\section{Introduction}

The \lhcb experiment~\cite{Alves:2008zz} has been conceived with the main goal
of studying heavy flavour physics in \pp collisions at the LHC, exploiting the unprecedented yield
of \bquark-quark pairs, which are mainly produced at small angles with respect to the direction
of the colliding proton beams. The detector is therefore designed as a forward spectrometer covering
the pseudorapidity region $2<\eta<5$ and providing excellent vertexing, tracking and
particle identification capabilities for the reconstruction of heavy flavour decays. 
Another key feature is the online selection system, consisting of a hardware level with high 
output bandwidth (up to 1 MHz), followed by a software level providing high flexibility.

Though heavy ion physics was not among the original motivations for the experiment, the detector
capabilities offer some unique possibilities also for this field:
\begin{itemize}
\item the forward acceptance, with a fully instrumented detector, highly complementary to the other LHC
experiments;
\item the excellent reconstruction performance, unrivaled at the LHC, for exclusive heavy flavour states 
down to null transverse momentum (\pt),
disentangling charmed particles produced promptly from those coming from \bquark-hadron decays;
\item the possibility to operate the detector in fixed target mode, by injecting small amount of noble
gas (He, Ne and Ar) in the LHC vacuum~\cite{smog} and studying beam-gas collisions, for which the forward geometry
of the detector is very well suited.
\end{itemize}

On the other hand, the most central \PbPb collisions 
can't be properly reconstructed due to the high track density in the forward region. \lhcb is therefore
more suited for smaller collision systems like \pPb, but can contribute also to study peripheral \PbPb collisions.
It has to be noted that studies in small collision systems have become increasingly important to the
interpretation of relativistic heavy ion collisions after the discovery, performed at the LHC, of sizeable 
collective behaviour even in \pp collisions~\cite{Khachatryan:2010gv}. 
 
\begin{figure}[tb]
  \centering
  \includegraphics[width=.9\textwidth]{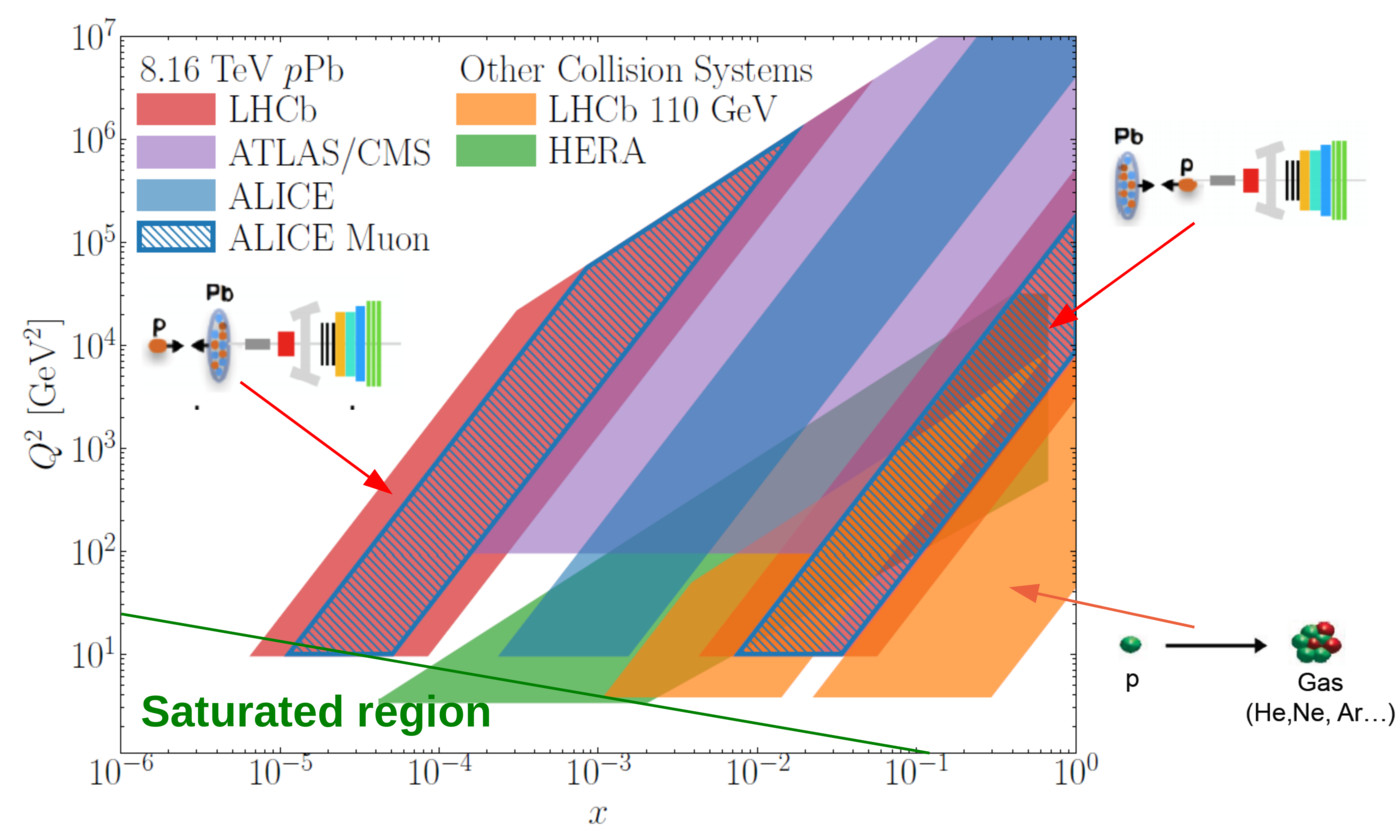}
  \caption{Kinematic reach corresponding to the acceptance of the four
    LHC experiments in \pPb collisions. The kinematic regions
    accessible in the fixed-target configuration at \lhcb and 
    in \electron-\proton collisions at HERA are also shown. }
  \label{fig:kinreach}
\end{figure}

A comparison among the kinematic reaches of \lhcb and the other LHC experiments in \pA collisions,
in terms of the Bjorken-$x$ value of the nucleon in the nuclear target and the squared 
parton-parton invariant mass $Q^2$, is shown in Figure~\ref{fig:kinreach}. 
Two regions are covered in \pPb collisions, depending on the orientation of the proton and lead beams. The 
so-called {\it forward} configuration, when the 
proton beam points toward the detector (i.e., it enters the detector region from its vertex detector), corresponds
to values of $x$ down to $10^{-5}$, where gluon saturation is expected to occur. In the {\it backward} configuration, 
when the Pb beam points toward the detector, measurements are
sensitive to the anti-shadowing region up to $x\sim 0.1$.

In fixed target collisions, the nucleon-nucleon centre-of-mass (c.m.)
energy reaches 110 GeV for proton beams of 6.5 TeV
energy. Measurements at this energy scale,  intermediate between
SPS fixed target experiments and beam-beam collisions at LHC,
provides an additional test bed to explore the energy evolution in the
dynamics of nuclear matter.
The detector acceptance corresponds to mid and backward rapidities in
the  c.m. frame: $-2.8 < y^* <0.2$. As also depicted in
Figure~\ref{fig:kinreach}, this gives access to large values of $x$
in the target nucleon, where nuclear PDFs are modified by the EMC
effect and where the contribution of a possible intrinsic heavy quark content in the
nucleon could be substantiated.

\section{\pPb collisions}

The experiment collected a first dataset of \pPb collisions at $\sqsnn = 5.02$ TeV in 2013,
corresponding to integrated luminosities of 1.1 and 0.5 \invnb in the forward and backward configuration,
respectively. After this succesfull experience, larger luminosities
were delivered to \lhcb during the 2016 \pPb run at  $\sqsnn = 8.16$
TeV, with 13.6 and 30.8 \invnb recorded. 
The results published so far by \lhcb focus on production of heavy flavour
states, whose modification with respect to \pp collisions
constitutes a major probe for the hot and dense matter, known as
quark-gluon plasma (QGP), produced in \PbPb collisions at LHC.
 In \pPb collisions  QGP is not expected to be formed  and the system
is considered to be a reference for the understanding of cold nuclear matter (CNM) effects.

\begin{figure}[tb]
  \centering
  \includegraphics[width=.32\textwidth]{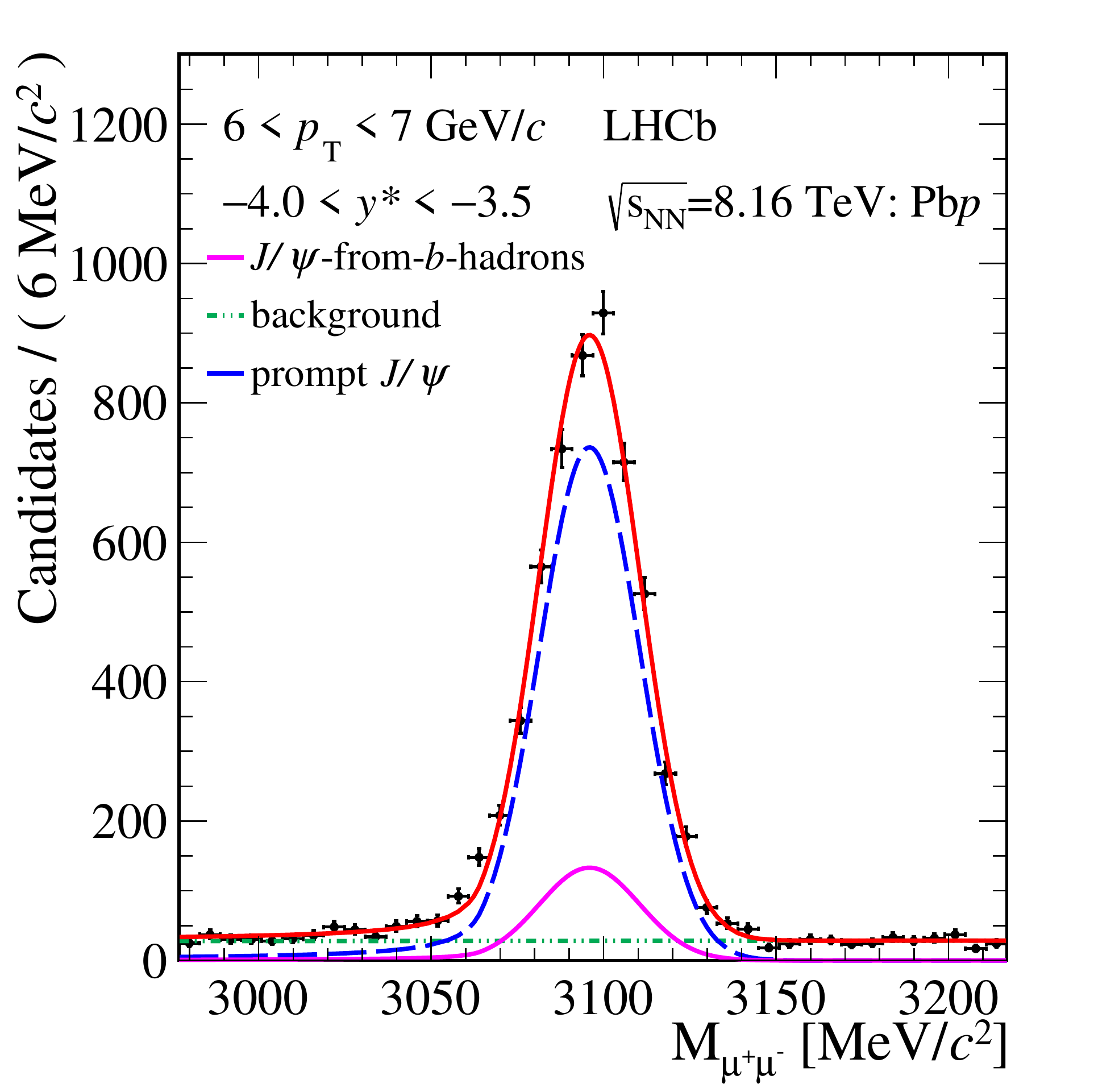}
  \includegraphics[width=.32\textwidth]{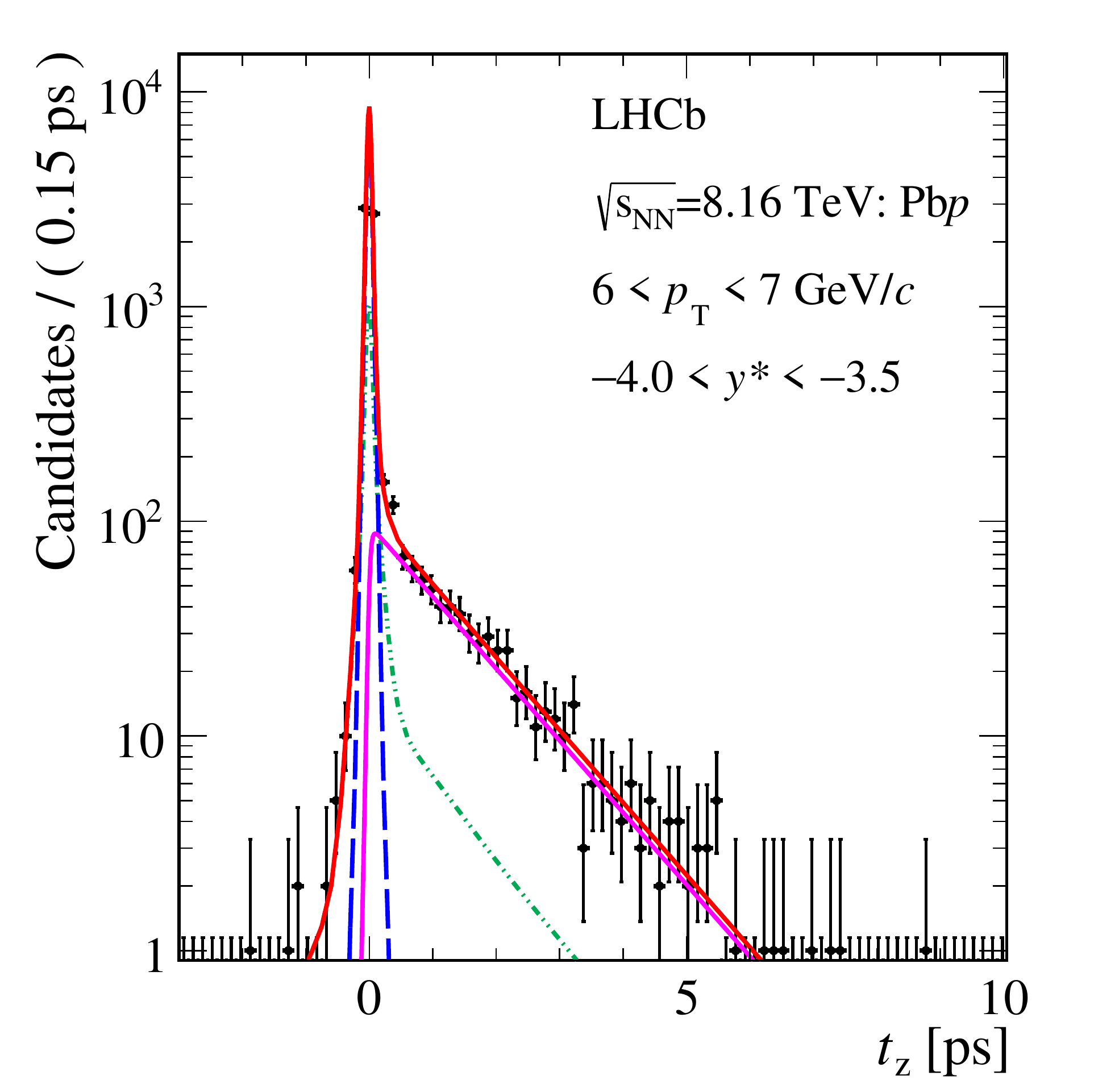}
  \includegraphics[width=.34\textwidth]{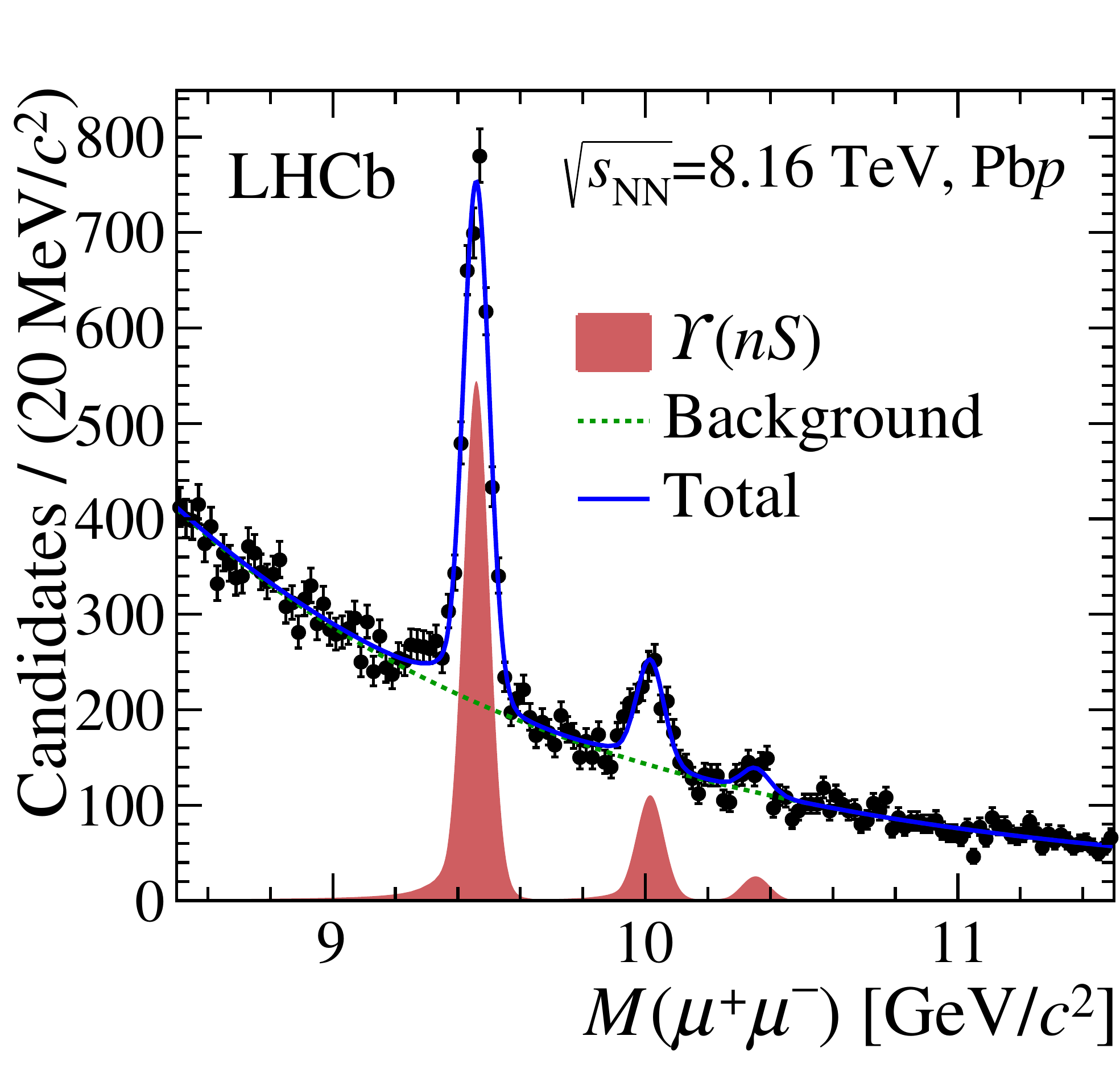}
  \caption{Distributions of (left) reconstructed mass and (middle) pseudo-proper time 
    \mbox{$t_z  \equiv (z_{\jpsi}-z_{PV}) \times (M/p_z)_{\jpsi}$} for the 
    $\jpsi\to\mumu$ decay candidates from the
    8 TeV \Pbp sample~\cite{LHCb-PAPER-2017-014} 
    in the rapidity bin $-4.0 < y^* <-3.5$ ($z_{\jpsi}$ and $z_{PV}$
    are the reconstructed longitudinal position of the \jpsi
    decay vertex and the interaction primary vertex, respectively). 
    The result of a fit to determine prompt signal,
    from-$b$ signal and background fractions is overdrawn.
    In the right plot. mass distribution for the 
    $\PUpsilon{(nS)}$ candidates from the same sample in
    $-5.0 < y^* <-2.5$~\cite{LHCb-PAPER-2018-035}.
}
  \label{fig:jpsireco}
\end{figure}

\begin{figure}[tb]
  \centering
  \begin{minipage}{.88\linewidth}
    \centering
  \includegraphics[width=.50\textwidth]{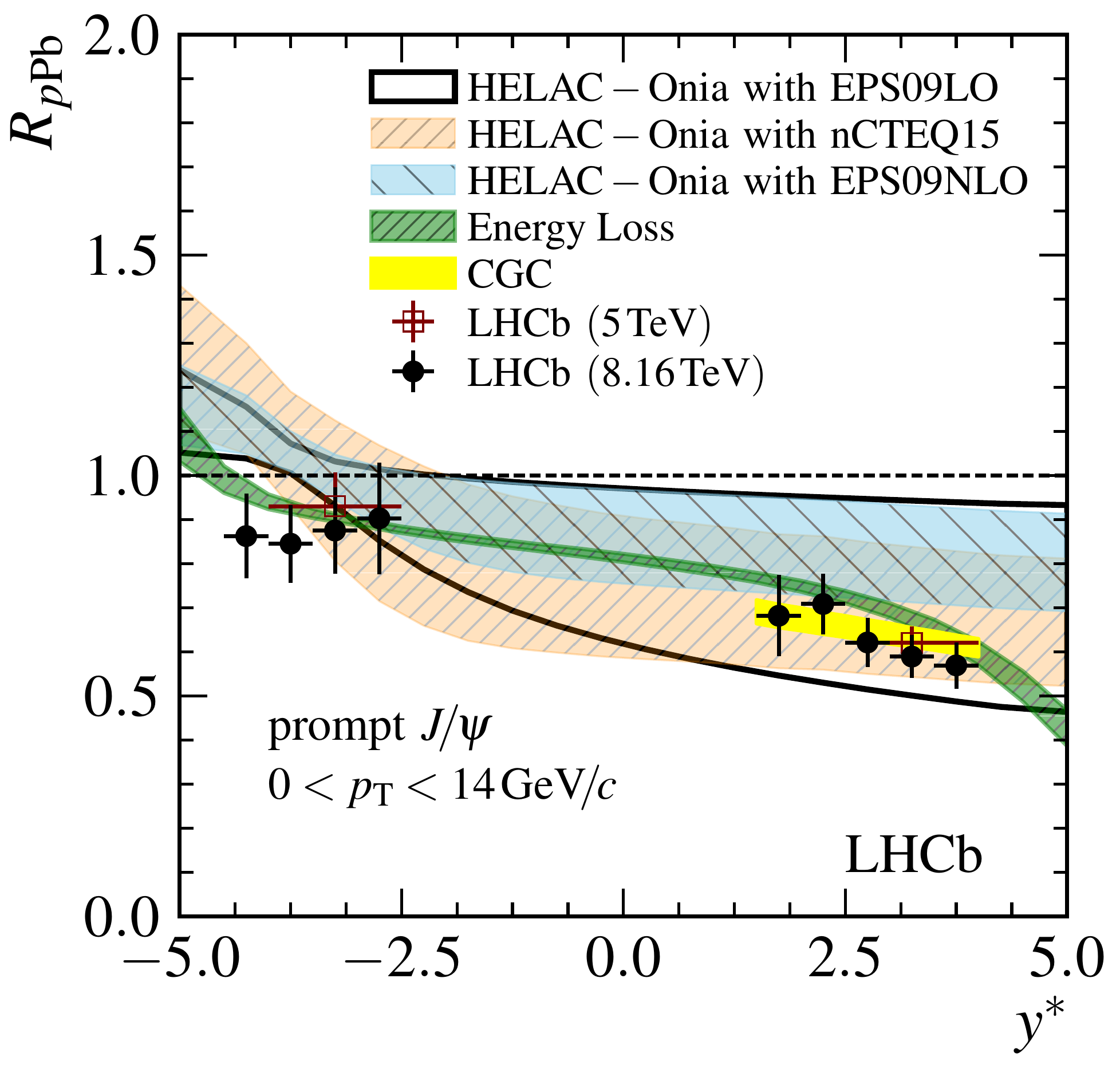}
  \includegraphics[width=.48\textwidth]{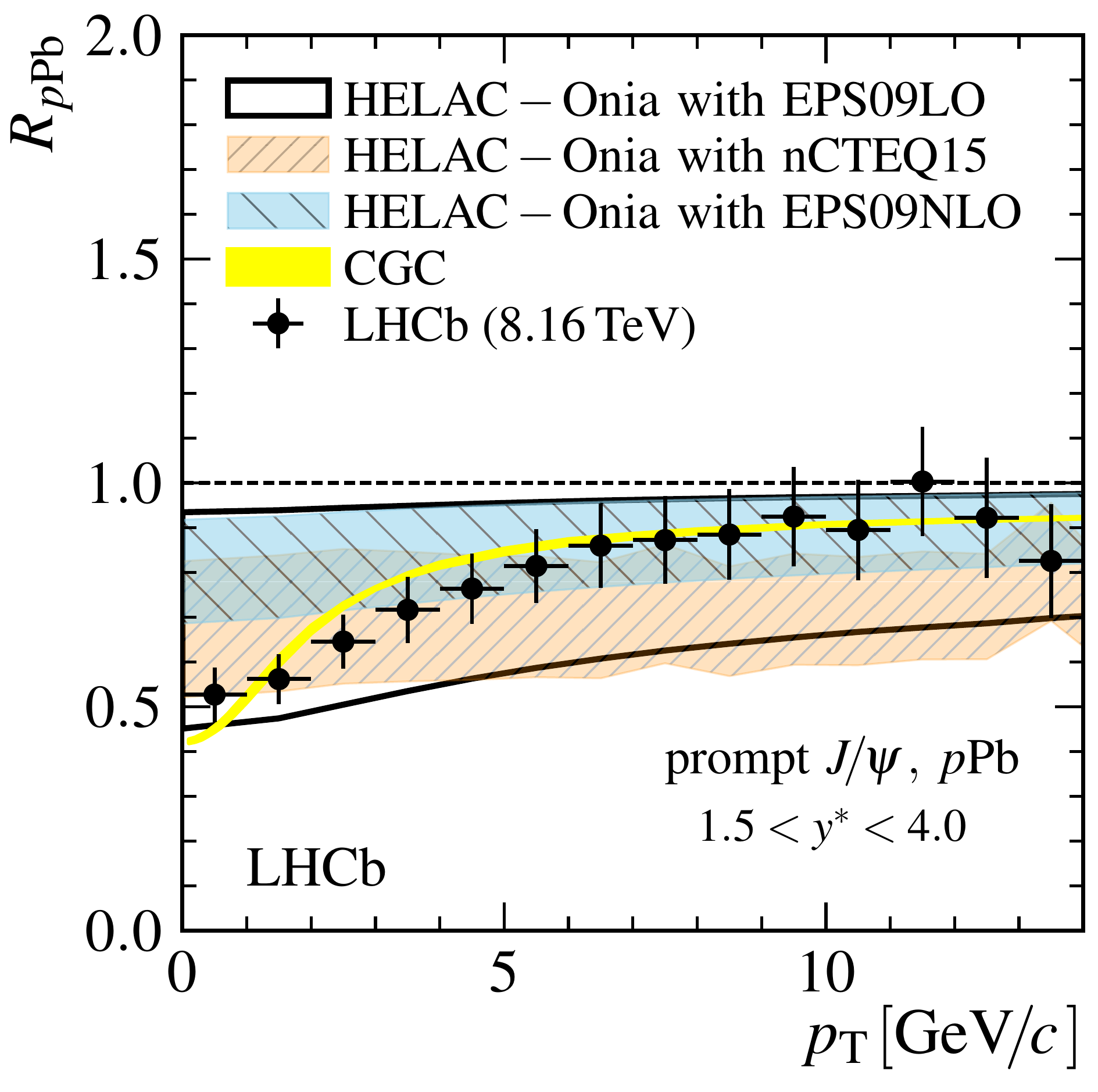}    
  \end{minipage}~~~
  \caption{Nuclear modification factor $R_{pPb}$ for prompt \jpsi
    production as a function of (left plot) $y^*$ and (right plot) \pt
    for the forward configuration~\cite{LHCb-PAPER-2017-014}. 
}
  \label{fig:jpsiresult}
\end{figure}

Figure~\ref{fig:jpsireco} illustrates the ability to distinguish
promptly produced $\jpsi$ mesons from those produced in $b$-hadron
decays in reconstructed $\jpsi\to\mumu$ decays.
The nuclear modification factor $R_{pPb}\equiv
\sigma_{p\text{Pb}}/(A_{\text{Pb}}~\sigma_{pp})$, where
$\sigma_{p\text{Pb}}$ and $\sigma_{pp}$ are the production
cross-sections in the two collision systems and $A_{\text{Pb}}=208$
is the lead mass number, can be measured separately for the two components.
The result for the prompt component obtained from the 8 TeV sample~\cite{LHCb-PAPER-2017-014}
is shown in Figure~\ref{fig:jpsiresult} as a function of \pt and the
c.m. rapidity $y^*$. A clear suppression with
respect to \pp collisions is observed at forward rapidity and low \pt,
compatible with the expected effect from nuclear PDF
modifications (shadowing) as computed in the framework of NRQCD
factorisation using several collinear nuclear PDF sets with the HELAC-Onia
package~\cite{Shao:2015vga,Lansberg:2016deg}.  
However, the rapidity dependence can also be
well explained by the coherent energy loss model~\cite{ArleoPeigne1303}.
The result at forward rapidity is also
compatible with the latest calculations based on the Color Glass Condensate model~\cite{Ducloue:2016pqr}.
More measurements, notably on Drell-Yan production, are needed to
disentangle these physical effects~\cite{Arleo:2015qiv}.

If final-state effects in CNM can't be excluded from the \jpsi result,
they are definetely needed to explain the different modifications of
quarkonia states. The first measurement of \psitwos production in the
5 TeV sample~\cite{LHCb-PAPER-2015-058} indicated a larger suppression with respect to \jpsi, in
agreement with results from the other LHC experiments~\cite{Abelev:2014zpa,Adam:2016ohd,Aaboud:2017cif,Sirunyan:2018pse}.
A more recent result~\cite{LHCb-PAPER-2018-035} shows evidence for different
suppression among the three $\PUpsilon{(nS)}$ states, which are cleanly observed
in the 8 TeV sample (see Figure~\ref{fig:jpsireco}).
\begin{figure}[tb]
\centering
 \begin{minipage}{.88\linewidth}
    \centering
  \includegraphics[width=.49\textwidth]{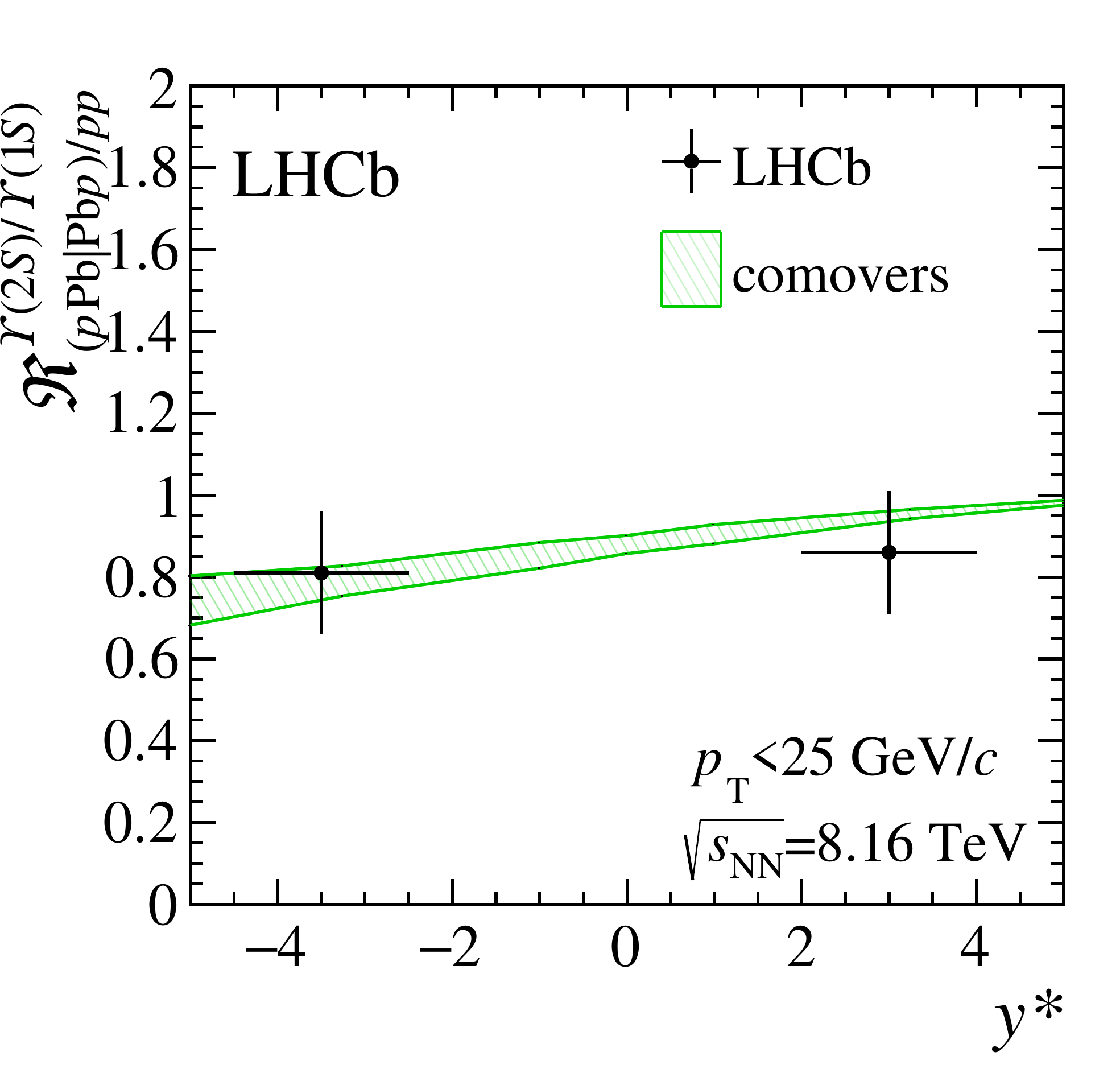}
  \includegraphics[width=.49\textwidth]{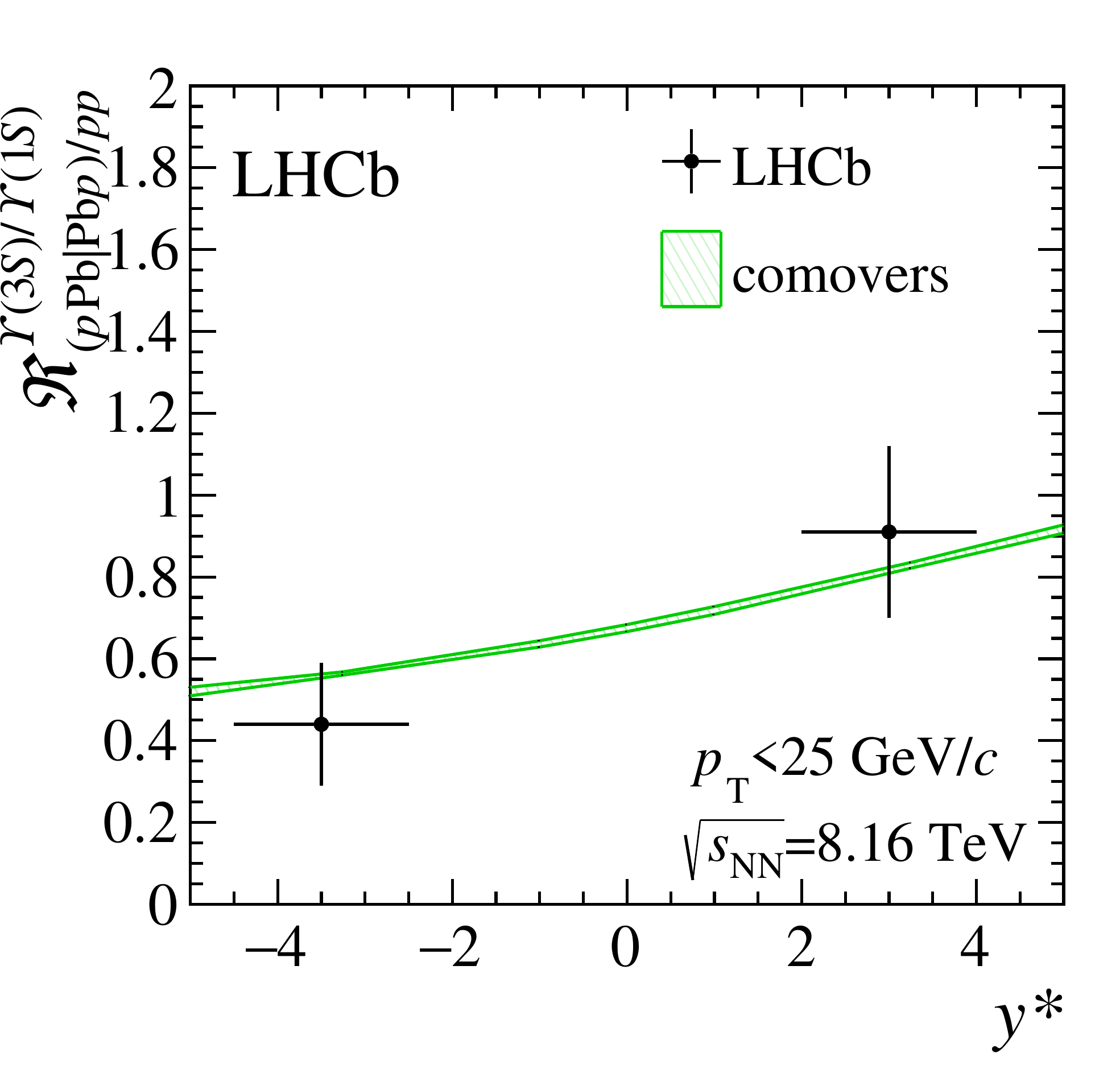}
  \end{minipage}
  \caption{Ratio between $\PUpsilon$ nuclear modification factors as a
    function of rapidity for (left plot) \Y2S/\Y1S and (right plot)  \Y3S/\Y1S~\cite{LHCb-PAPER-2018-035}.
}
  \label{fig:YpPb}
\end{figure}
The result, shown in Figure~\ref{fig:YpPb},  is nicely described
in the framework of the ``comovers'' model~\cite{Ferreiro:2018wbd}, 
where dissociation of the quarkonia states is attributed to
interaction with final-state particles which are close in phase-space.
This relatively large effect needs to be taken into account in the
interpretation of the spectacular suppression of $\Upsilon$
states recently observed by CMS in \PbPb collisions~\cite{Sirunyan:2018nsz}.

\begin{figure}[tb]
  \centering
  \includegraphics[width=.49\textwidth]{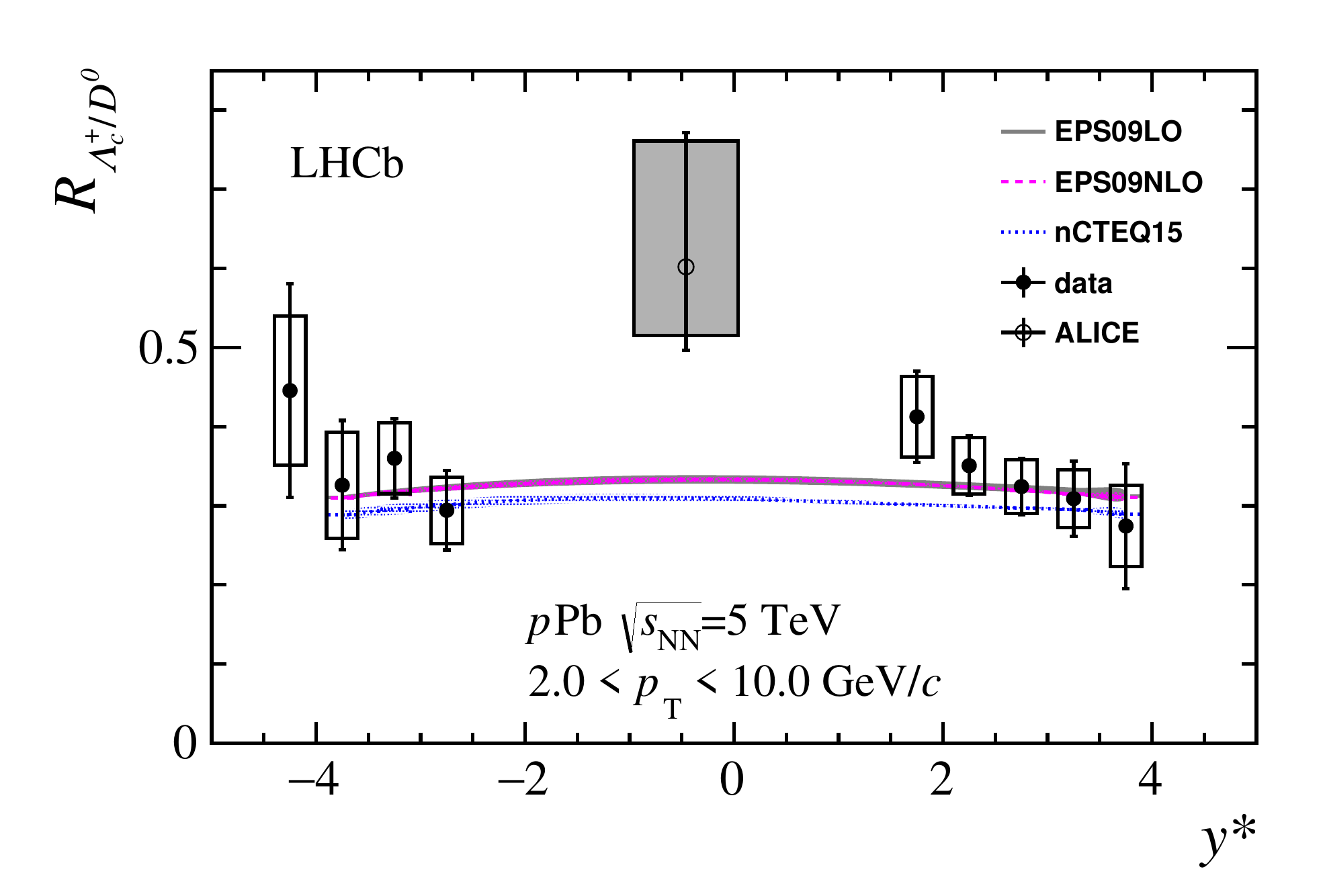}  
  \includegraphics[width=.49\textwidth]{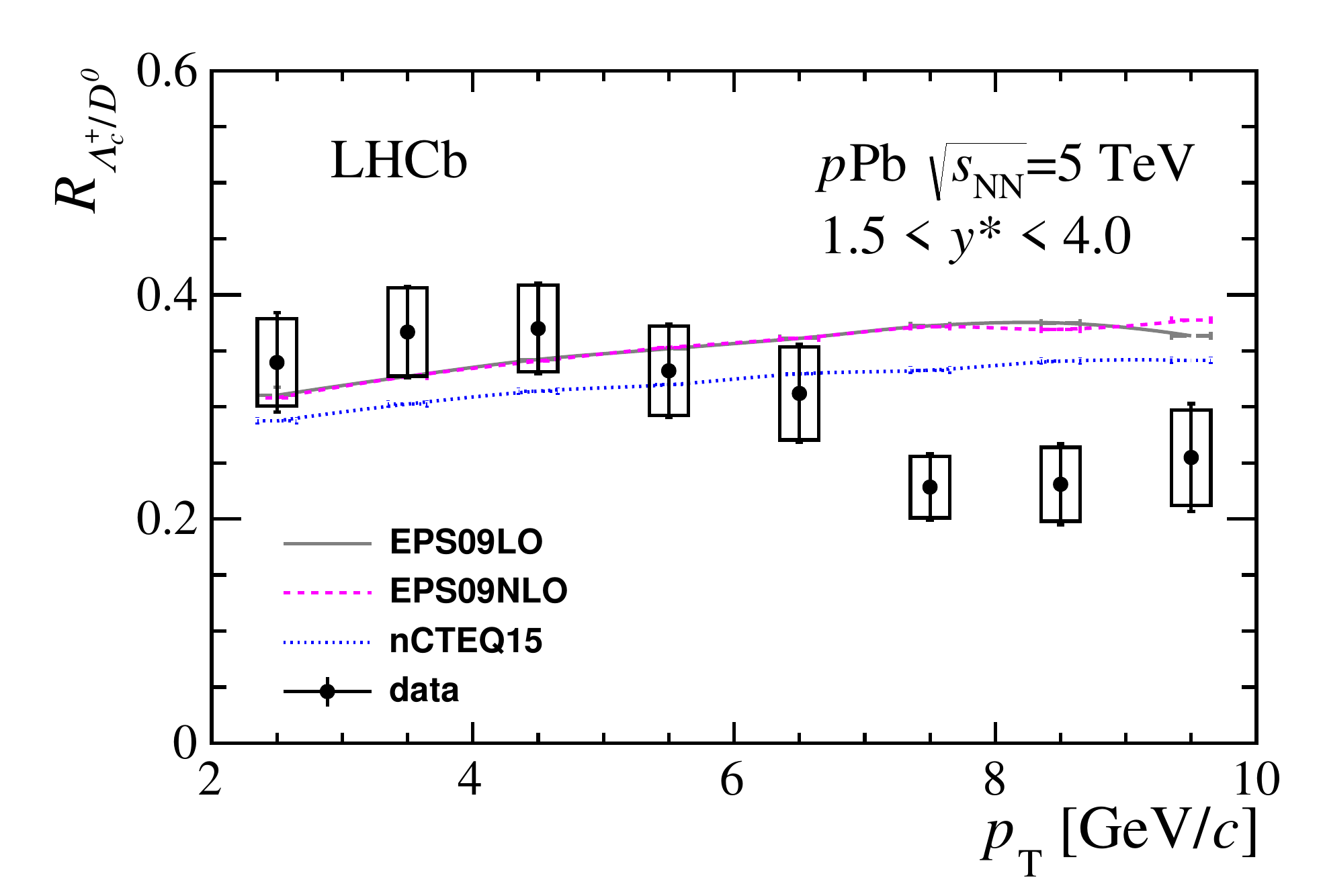}
  \caption{Ratio between \Lc and \Dz production in \pPb collisions at
    5 TeV, as a function of (left plot) $y^*$ and (right plot) \pt
    for the forward configuration~\cite{LHCb-PAPER-2018-021}. 
    The result from ALICE~\cite{Acharya:2017kfy} at mid
    rapidities is also shown for comparison.
  }
  \label{fig:Lcratio}
\end{figure}

Production of open charm states, \Dz mesons~\cite{LHCb-PAPER-2017-015} 
and \Lc baryons~\cite{LHCb-PAPER-2018-021}, has also been
measured  with large statistics already in the 5 TeV sample.
Data are precise enough to constrain nuclear PDFs, assuming that
initial-state effects dominate the observed nuclear modification.
The \Lc/\Dz ratio is an important input to the hadronisation
phenomenology, since baryon enhancement in heavy ion collisions, 
notably at low \pt, is expected from production via coalescence and is
affected by the thermal properties of the nuclear medium. 
\lhcb measures a ratio between the \Lc and \Dz prompt production
around 0.3, with no evidence of strong dependence on 
the rapidity or transverse momentum (see Figure~\ref{fig:Lcratio}). Results
are in substantial agreement with HELAC-onia computations where the
only nuclear effects are due to PDF modifications and 
largely cancel in the ratio.

\begin{figure}[tb]
  \centering
  \begin{minipage}{.9\linewidth}
    \centering
  \includegraphics[width=.49\textwidth]{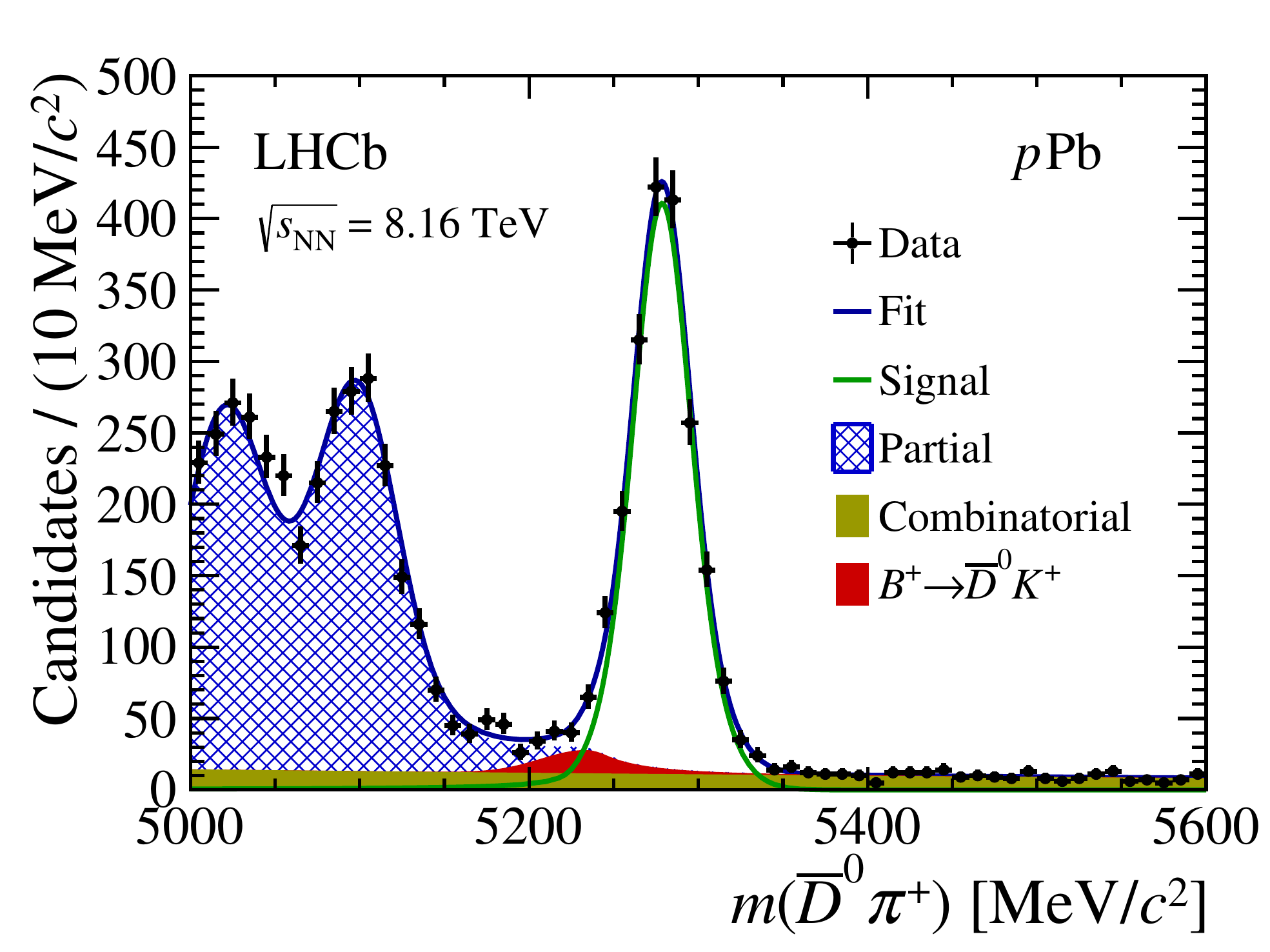}
  \includegraphics[width=.49\textwidth]{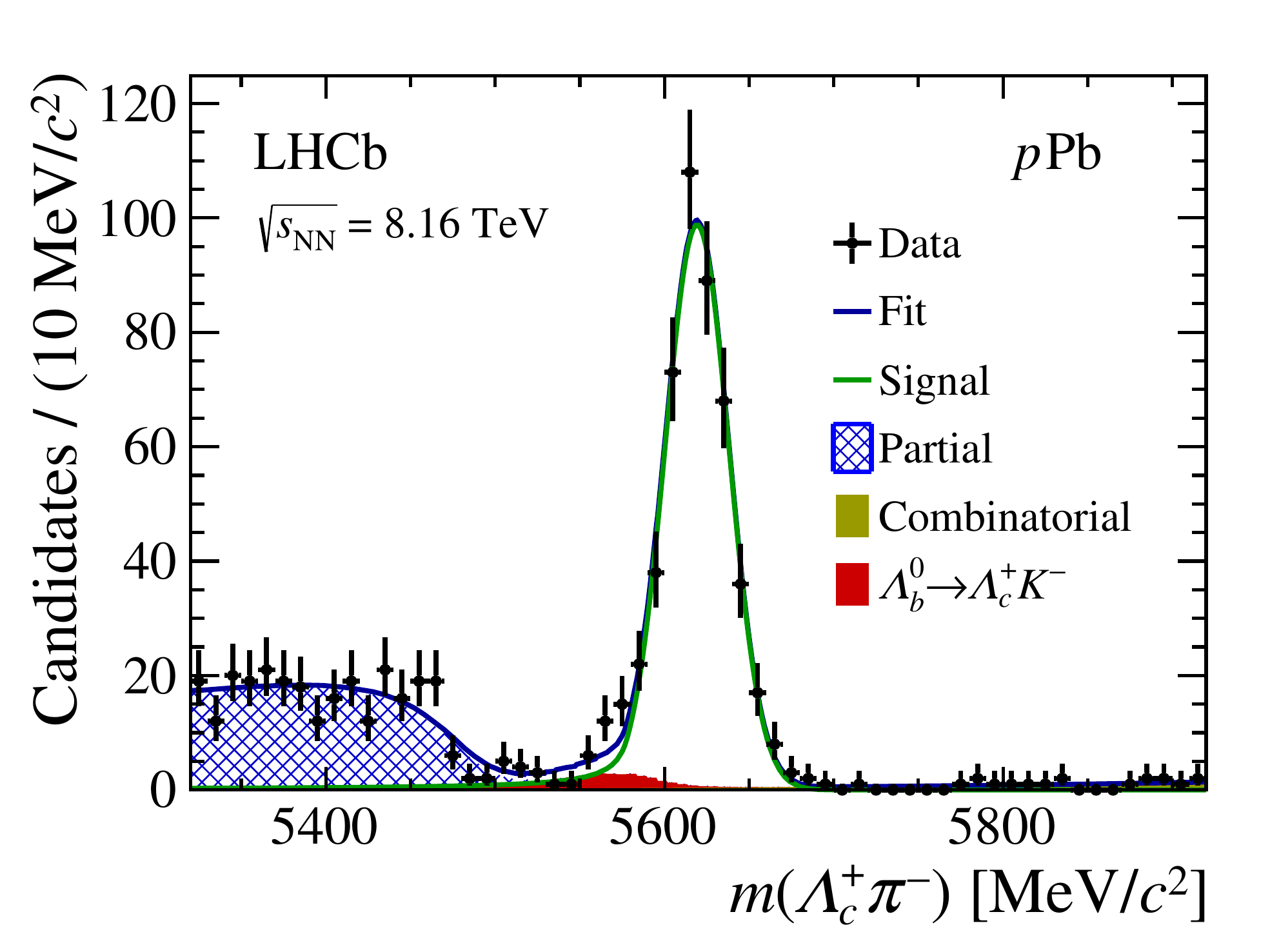}    
\end{minipage}
\begin{minipage}{.88\linewidth}
  \centering
  \includegraphics[width=.49\textwidth]{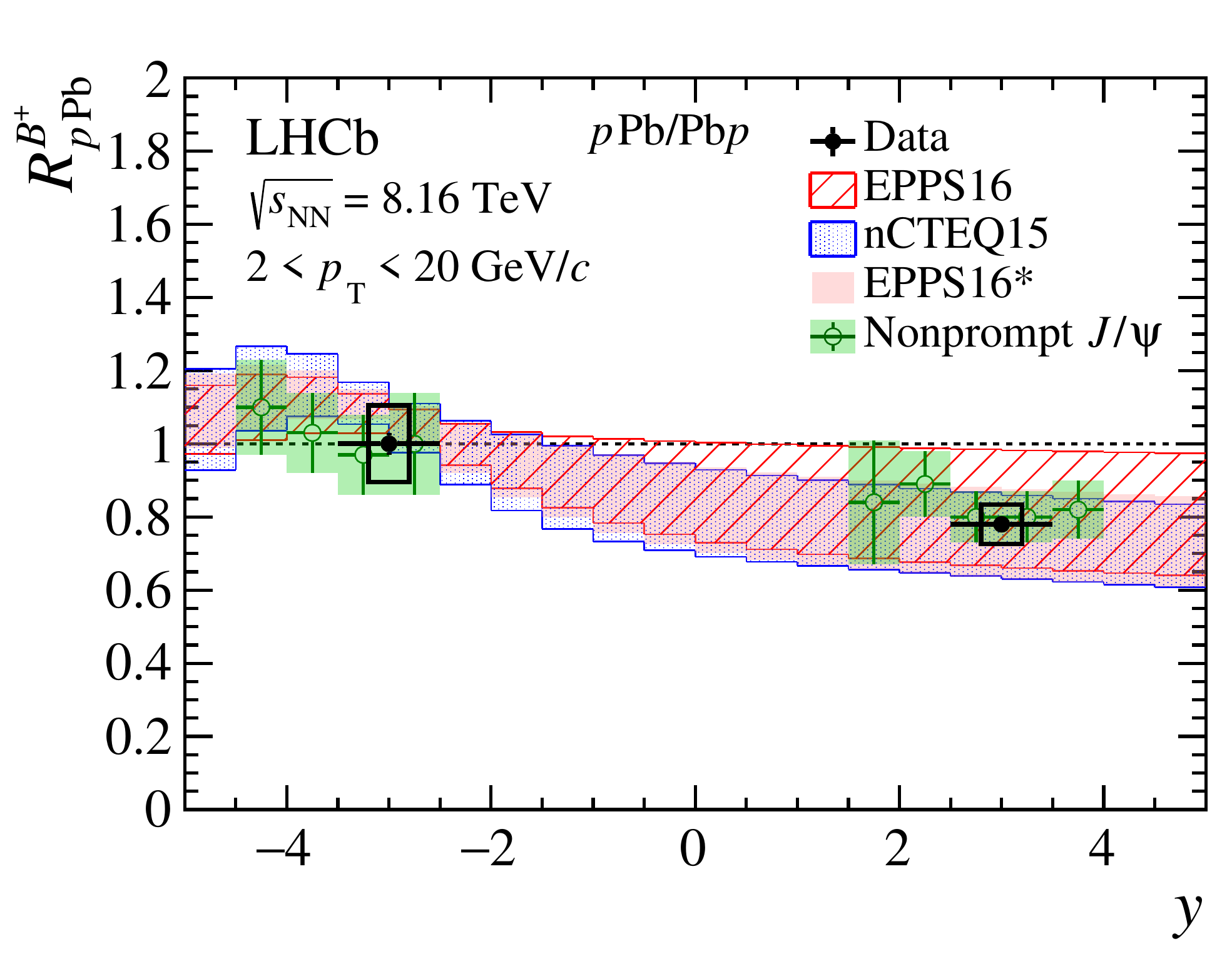}
  \includegraphics[width=.49\textwidth]{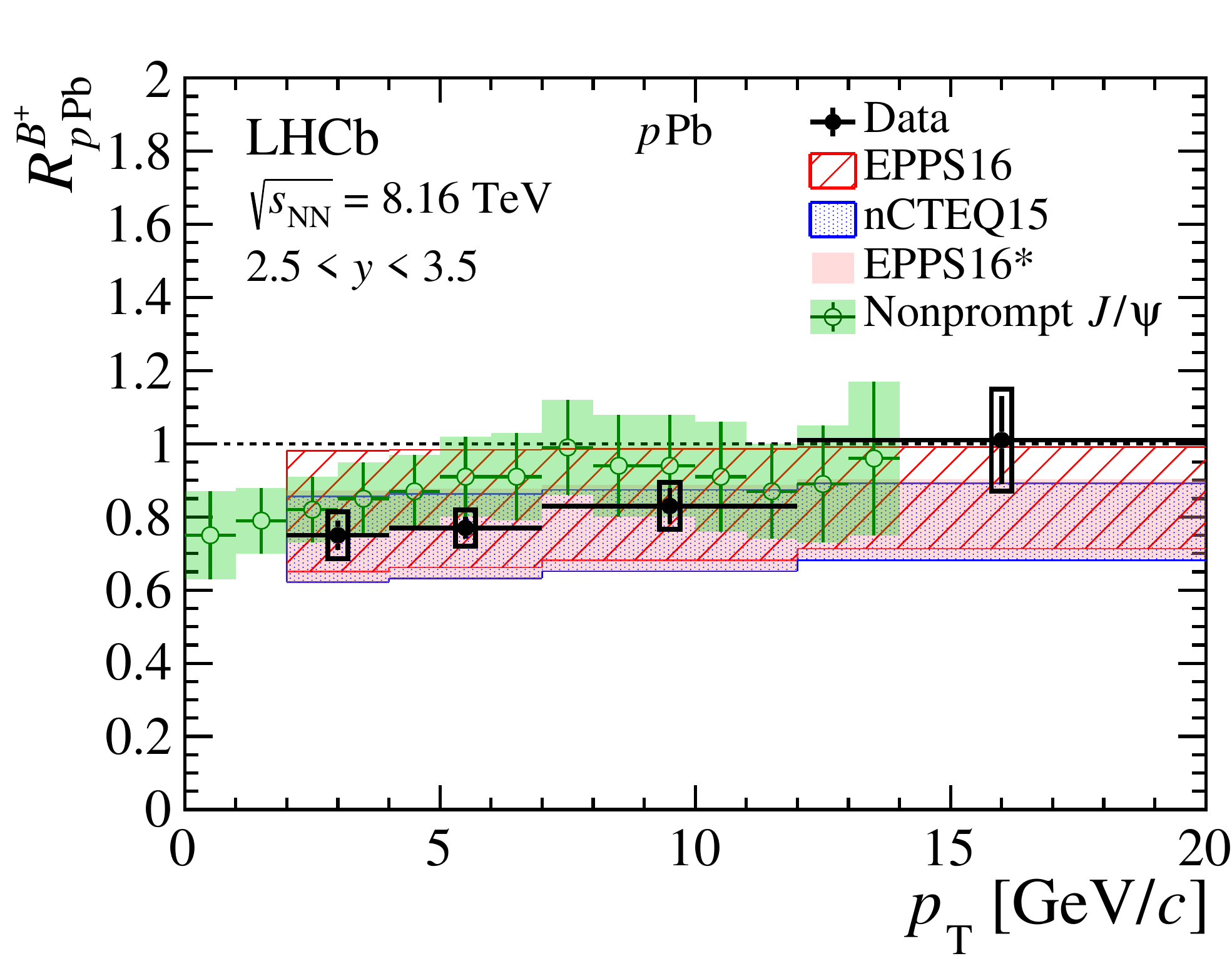}
\end{minipage}
  \caption{In the upper plots, reconstructed mass distribution of
    candidates for the (left) $\Bp\to\Dzb\pip$ and (right)
    $\Lb\to\Lc\pim$ decays in the \pPb forward configuration data at 8
    TeV. In the lower plots, the resulting nuclear modification factor $R_{pPb}$ for \Bp meson
    production  is plotted as a function of 
    (left plot) $y^*$ and (right plot) \pt
    for the forward configuration~\cite{LHCb-PAPER-2018-048}.
    Results are compared with the measurement for \jpsi from $b$-hadron 
    decays~\cite{LHCb-PAPER-2017-014} and with HELAC-onia calculations
    using different nuclear PDF sets. 
  }
  \label{fig:openB}
\end{figure}

For the first time, exclusive $b$-hadron decays
have been reconstructed in nuclear collisions~\cite{LHCb-PAPER-2018-048}.
Clean samples of four decay modes ($\Bp\to\Dzb\pip, \Bp\to \jpsi\Kp,
\Bz\to\Dm\pip, \Lb\to\Lc\pim$) have been obtained from the 8 TeV \pPb
dataset, down to \pt values below the hadron mass.
The \Bp modes  have been used to measure the nuclear modification
factor as a function of $y^*$ and \pt. The results, shown in
Figure~\ref{fig:openB}, confirm the suppression pattern observed with
detached \jpsi, consistently with nuclear PDF effects. The ratio
between prompt production of \Lb baryons and \Bz mesons is also
measured and found consistent with the \pp measurement.

\section{\PbPb collisions}

A first small sample of \PbPb collisions was recorded by \lhcb during the
2015 run, corresponding to about $10~\mu\text{b}^{-1}$. The
tracking detector performance in this challenging environment was found to be
satisfactory for events of centrality above 50\%. 

\begin{figure}[tb]
  \centering
  \includegraphics[width=.7\textwidth]{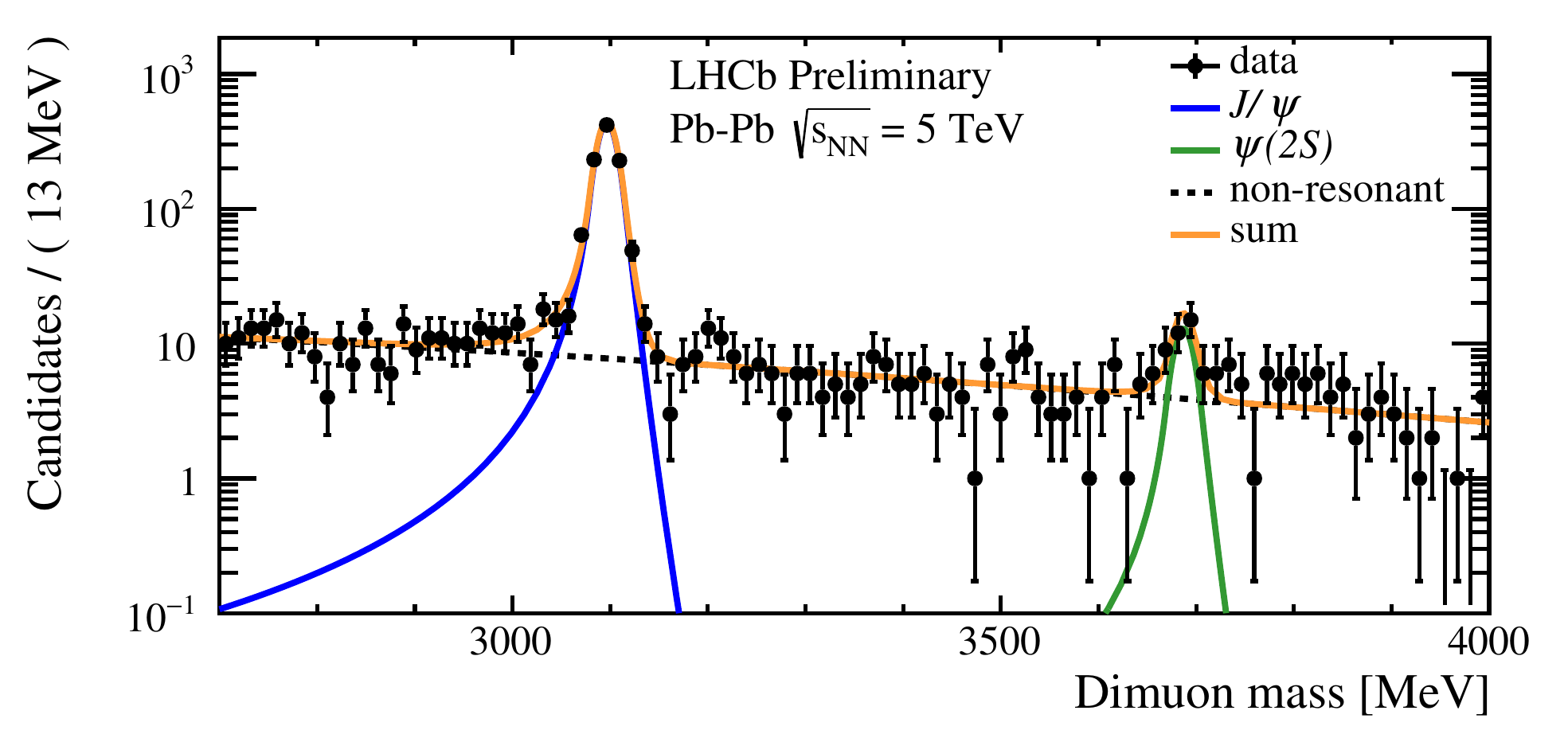}
  \includegraphics[width=.7\textwidth]{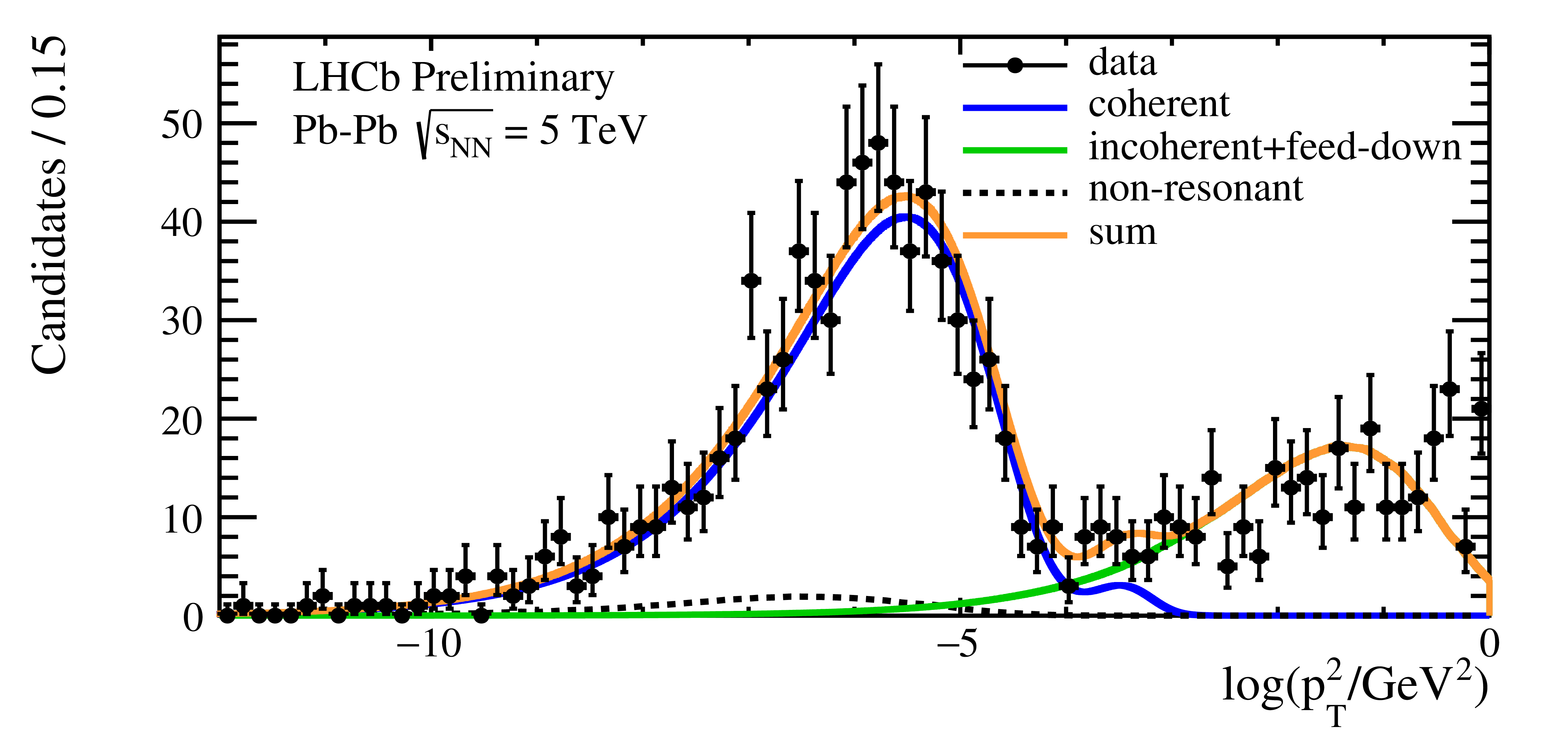}
  \caption{ Candidates for \jpsi exclusive production
    in \PbPb collisions at 5 TeV~\cite{LHCb-CONF-2018-003},
    reconstructed from $\jpsi\to\mumu$ decays.
    In the upper plot, the reconstructed mass distribution is shown in an
    extended mass range where, beside the clean \jpsi peak, a small signal for 
    \psitwos can also be seen. In the lower plot, the $\log(\pt^2)$ distribution
    in the \jpsi mass range ($3096.9 \pm 65~\mev$) is fitted with templates for coherent and
    incoherent production.
  }
  \label{fig:upcjpsi}
\end{figure}

The first preliminary physics result~\cite{LHCb-CONF-2018-003} has been obtained from ultraperipheral
collisions (UPC), where hadron photoproduction is enhanced by the 
large photon flux from the lead nuclei. The observation of photoproduction of heavy
flavour states, providing a hard scale for perturbative QCD
calculations,  is particularly interesting to explore the gluon density
down to the saturation region at $x\sim 10^{-5}$.
The exclusive production of \jpsi is cleanly
observed (see Figure~\ref{fig:upcjpsi}). The excellent \pt resolution
allows to distinguish coherent and uncoherent production, whose \pt
distributions are found to be well described by templates obtained with the
STARlight generator~\cite{Klein:2016yzr}. 

The accuracy of the result is limited by the size of the data sample,
but this analysis demonstrates the \lhcb potential for physics in
\PbPb UPC.
During the run performed in november 2018, an integrated
luminosity of $210~\mu\text{b}^{-1}$ was collected, providing
the possibility for a precision measurement of exclusive \jpsi and
\psitwos production, and possibly to access rarer channels like
$\PUpsilon$ production and light-by-light scattering.

\section{Fixed-target collisions}

\begin{figure}[b]
  \centering
\includegraphics[width=.96\linewidth]{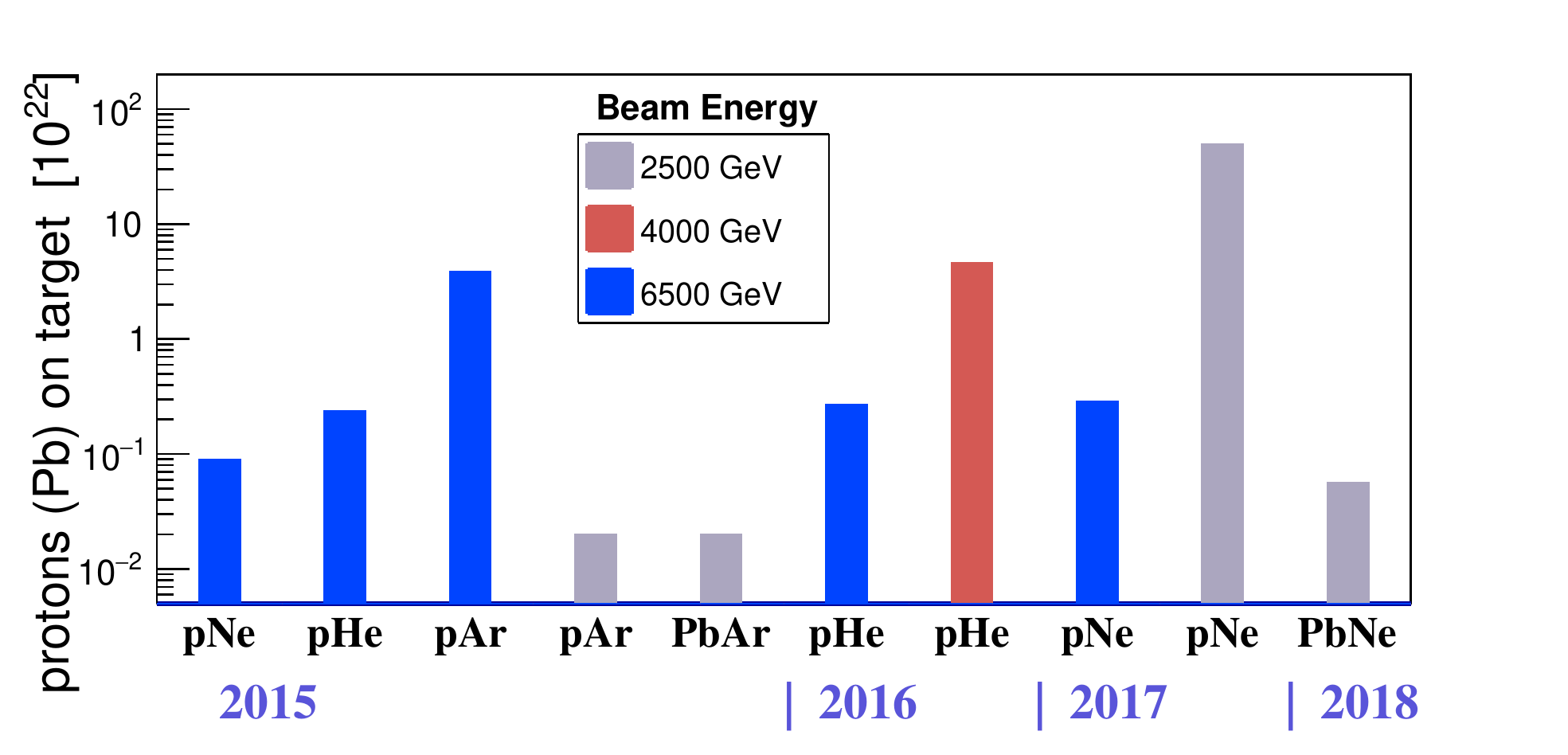}
  \caption{Summary of collected fixed-target physics samples at
    \lhcb. 
The data size is given in terms of delivered protons (ions) on target
(POT). For a nominal target pressure 
of $2\times 10^{-7}$\,mbar, $10^{22}$\,POT correspond to an integrated luminosity of 
about $5$\,nb$^{-1}$ per meter of gas, though the actual target pressure
and the data taking efficiency vary among samples.  }
  \label{fig:smogsamples}
\end{figure}

Samples of beam-gas collisions with proton and lead beams of different
energy, and with the three possible targets (He, Ne and Ar) have been
collected by \lhcb during the LHC Run 2, as summarized in
Figure~\ref{fig:smogsamples}.  
The largest sample, \pNe collisions at $\sqsnn=69~\gev$, corresponds
to an integrated luminosity of about 100~\invnb. A sample of Pb-Ne
collisions at the same energy has been recently collected during the
2018 \PbPb run.  

\begin{figure}[tb]
  \centering
  \includegraphics[width=.49\textwidth]{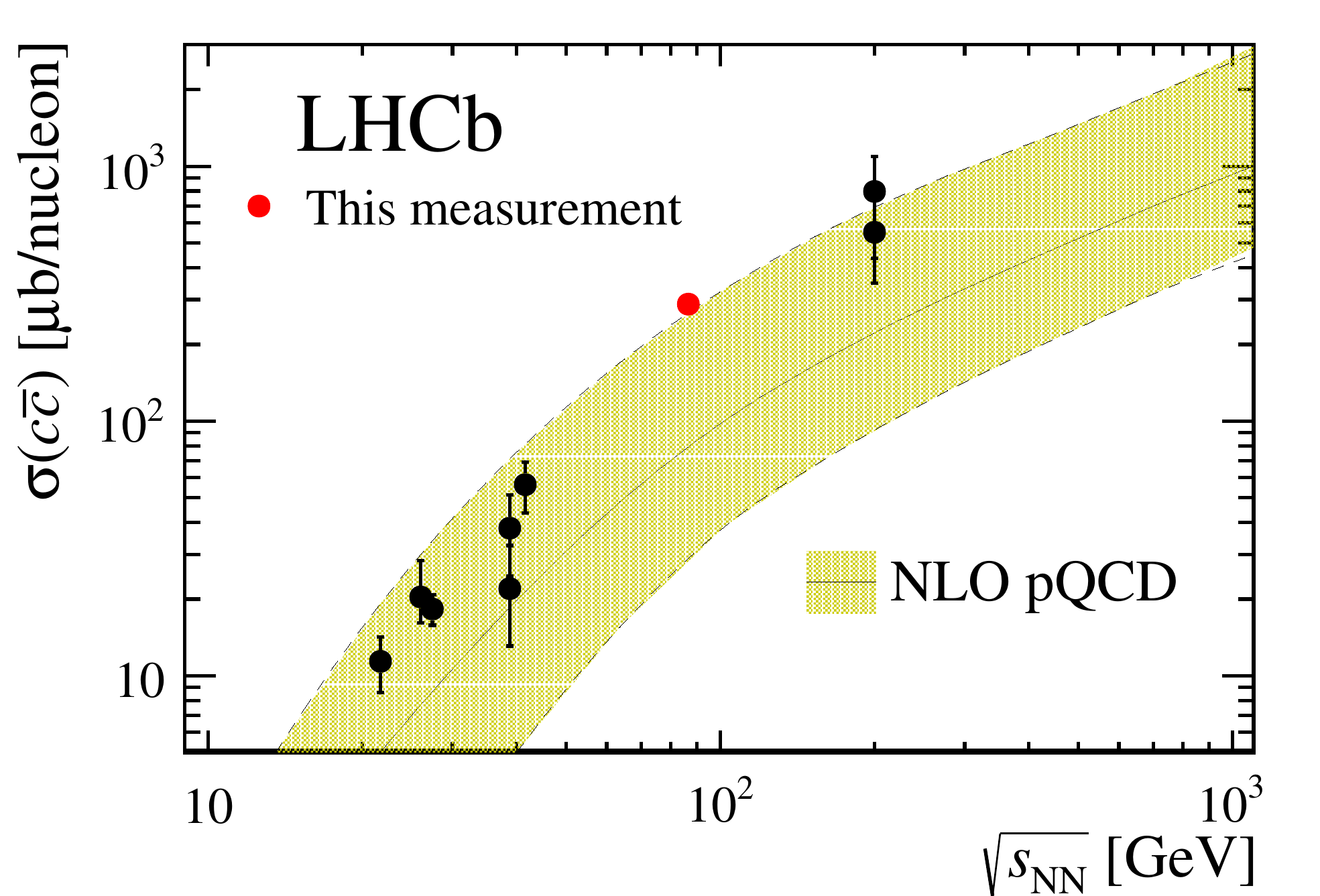}
  \includegraphics[width=.49\textwidth]{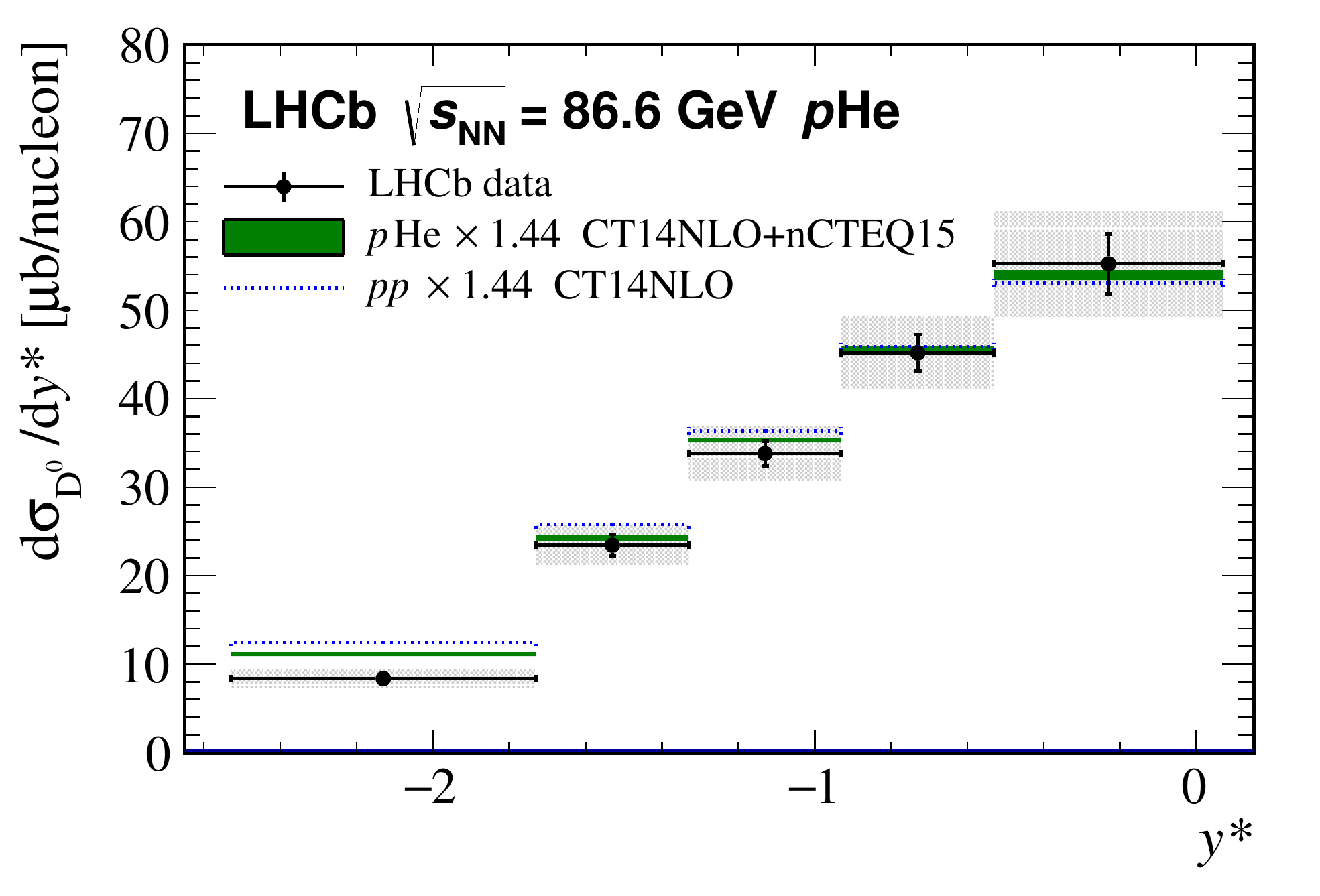}
  \caption{ In the left plot, the result for the \ccbar production
    cross-section per nucleon, obtained from \pHe collisions at
    $69~\gev$~\cite{LHCb-PAPER-2018-023}, is compared with other
    experimental results and with a NRQCD
    calculation~\cite{Maltoni:2006yp}. The right plot  
    shows the rapidity dependence of the \Dz
    production cross-section in the same sample. The
    result is compared with HELAC-onia calculations, not accounting
    for intrinsic charm, which are rescaled to reproduce the
    integrated value of the measured cross-section.    
  }
  \label{fig:charmFT}
\end{figure}

The first physics results, demonstrating the potential of this novel
program, have been obtained from some of the first samples collected in 2015
ane 2016. The first measurement of charm production~\cite{LHCb-PAPER-2018-023}
is based on the $7.6 \pm 0.5~\invnb$ of \pHe collisions at $87~\gev$,
and a few \invnb of \pAr data at $110~\gev$. Order of $10^3$ \Dz mesons
and $10^2$ \jpsi mesons are reconstructed from both samples. These data
provide the first determination of the \ccbar cross-section at this
relatively unexplored energy scale (see Figure~\ref{fig:charmFT}). The rapidity dependence of the
production is found to agree with predictions based on collinear
factorisation not including a contribution from intrinsic charm, as
shown in Figure~\ref{fig:charmFT}.
Therefore, these measurements, which are sensitive to $x$ values up to
$\sim 0.5$, do not favour the large intrinsic charm contributions  
predicted by some models in this kinematic range\cite{Pumplin:2007wg,Brodsky:2015fna}.
The larger available datasets, and possibly the use of a hydrogen
target in the future, will allow for more accurate constraints.

\begin{figure}[tb]
  \centering
  \includegraphics[width=.54\textwidth]{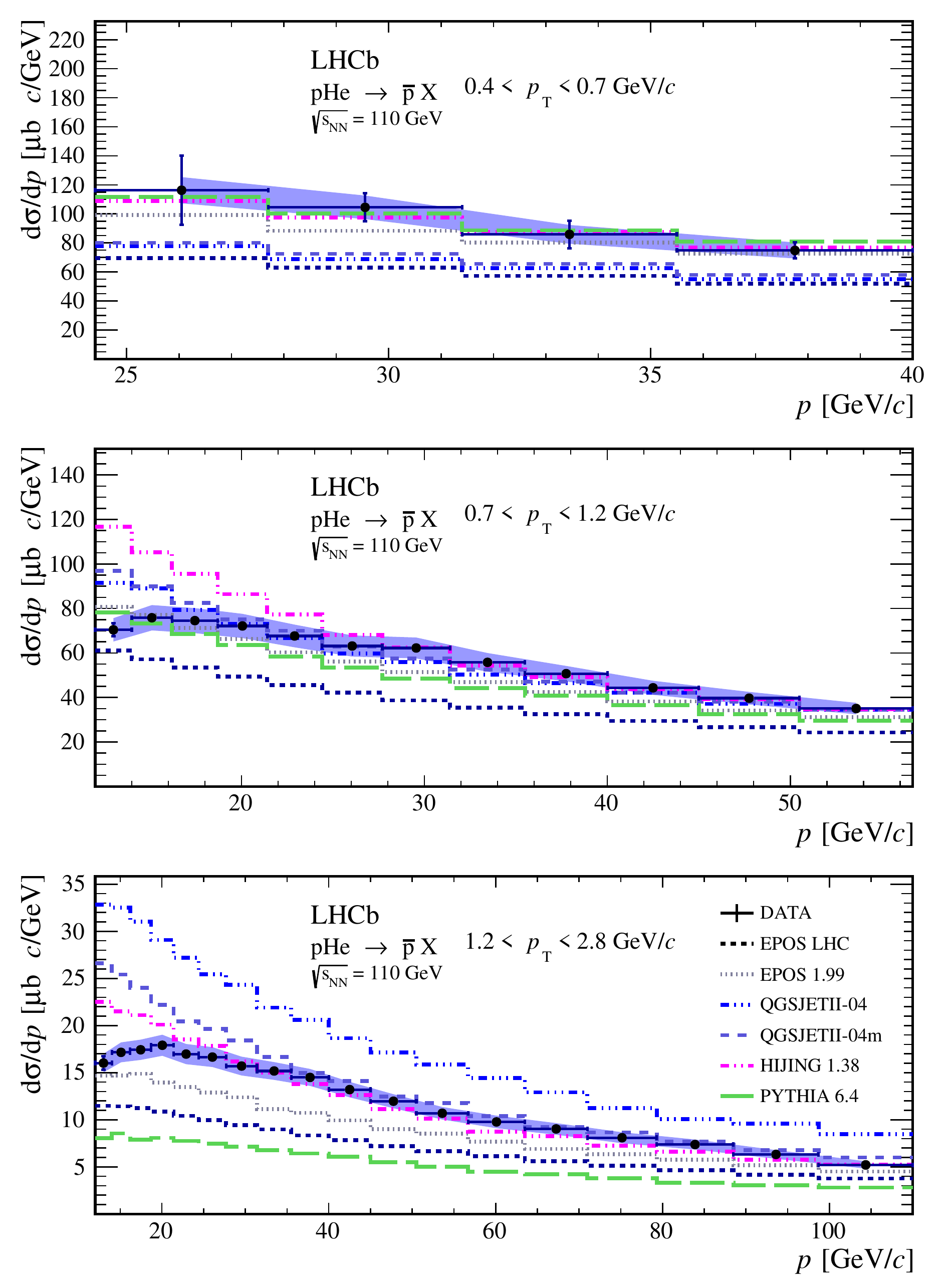}
  \includegraphics[width=.45\textwidth]{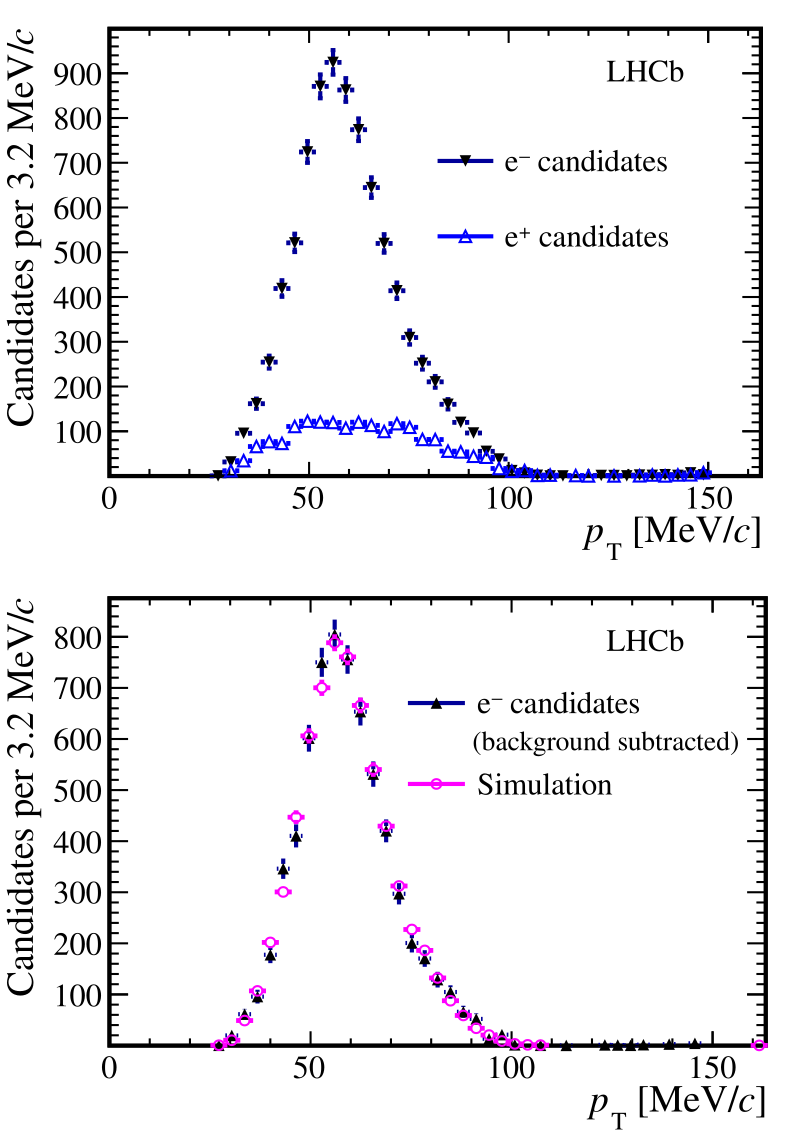}
  \caption{On the left plot, result for antiproton production in \pHe
    collisions at $110~\gev$. The differential production
    cross-section $d\sigma/dp$ is shown as a function of the momentum
    \ptot for different ranges of \pt and is compared with several
    generators included in the CRMC package~\cite{crmc}. The
    normalization is obtained from the {\pe}elastic scattering events
    from the same sample, illustrated in the  right plots.
    The upper plot shows the \pt distribution for single scattered electron
    candidates with negative and positive charge, the latter sample
    being used to subtract the background from hadronic collisions. 
    After subtraction, the
    distribution is found to agree well with the expected one
    from simulated {\pe}scattering, as shown in the lower plot.
  }
  \label{fig:phepbar}
\end{figure}

The antiproton production in the \pHe sample at $110~\gev$~\cite{LHCb-PAPER-2018-031} has also been
measured. This study is motivated by the recent precision measurements performed
in space, notably by AMS-02~\cite{AMS02}, of the antiproton content in cosmic rays,
which is sensitive to possible exotic contributions like dark matter
annihilation. For antiprotons above 10 GeV, the largest uncertainty on
the expected flux of antiprotons from known sources, namely production in collisions
between primary cosmic rays and the interstellar medium, is due to the
limited knowledge of the corresponding production
cross-sections. \lhcb performed the first \pbar production measurement
in \pHe collisions, which are responsible for about 40\% of the
expected cosmic \pbar flux, in the range of \pbar momentum between $12$
and $110~\gev$. The results, shown in
Figure~\ref{fig:phepbar}, are significantly more precise than the
spread among the predictions from different phenomenological models,
and are contributing to improve the models for cosmic secondary \pbar
and the resulting constraints on dark matter 
contributions~\cite{Reinert:2017aga,Korsmeier:2018gcy}.

For both results, the integrated luminosity of the sample is
estimated from the yield of elastically scattered electrons
from the target atoms. The yield is determined by selecting events with a single
low-pt track in the detector, identified as an $e^{\pm}$. The
background to this normalization channel is
due to soft diffractive collisions with a single track reconstructed
in the detector and is expected to be symmetric in charge, so that it can
be estimated and subtracted using the $e^+$ candidates, 
as illustrated in Figure~\ref{fig:phepbar}. The integrated luminosity
is determined using this method with an accuracy of
6\%~\cite{LHCb-PAPER-2018-031}, dominated by the systematic
uncertainty on the electron reconstruction efficiency.

\section{Conclusions and Prospects}

The results obtained so far by the \lhcb collaboration from heavy ion
collisions demonstrate the capability of the experiment to provide
unique contributions to this field, exploiting a variety of
collision systems and energy scales, where exclusive final states,
notably in the heavy flavour sector, can be reconstructed with high
efficiency and purity. 

The rich datasets collected
so far provide many additional possibilities that are being investigated. In
\pPb collisions, studies on direct photons, both inclusive and in
gamma-jet events, are underway, with unique sensitivity to the
saturation region~\cite{CesarPoster}. The size of the sample collected
in 2018 at 8 TeV is expected to provide access to new channels, as
\chic and \etac quarkonia states and Drell-Yan dimuon events. 
Studies of flow and correlations with identified particles in the
unique forward acceptance region covered by \lhcb are also planned.

Substantial development of the heavy ion program in \lhcb is also
expected during the future LHC runs, taking profit of the currently
ongoing detector upgrade~\cite{LHCb-TDR-012} which includes a new vertex
detector with improved granularity. The proposed plans for future
heavy ion running, discussed in detail in Reference~\cite{Citron:2018lsq}, 
foresee an increase in integrated luminosity for \pPb and \PbPb collisions
by more than an order of magnitude during the LHC Runs 3 and 4
(2021-2029). This will open novel possibilities, as precision studies
using Drell-Yan events and correlations in heavy flavour production.
A proposal for a second detector upgrade for the LHC Run 5 (starting
2031) has been put forward~\cite{LHCb-PII-EoI,LHCb-PII-Physics}. 
The granularity of such detector,
conceived to take profit of the full potential of LHC luminosity in \pp collisions, would
make it possible to reconstruct even the most central \PbPb collisions.

The fixed-target program will also be considerably developed, thanks
to the installation of a new target device~\cite{DiNezza:2651269}. 
This will consist of a storage cell containing the injected gas into
a 20 cm long region located  just upstream the \lhcb vertex detector. 
An increase in fixed-target luminosity by up to two orders of magnitude with the same injected gas flow
is anticipated. Furthermore, it will be possible to inject more gas
species, notably hydrogen, providing \pp reference for the other fixed-target
samples.  The physics potential of such program,
discussed in more detail in Reference~\cite{Bursche:2649878}, covers a
wealth of measurements that won't be possible at any other facility in
the coming years, including heavy flavour and Drell-Yan production
giving access to nuclear PDFs at large $x$, measurements of 
quarkonia suppression in different systems at $\sqsnn\sim 100~\gev$,
and studies of particle production of great interest for cosmic
ray physics.

\section*{References}
\setboolean{inbibliography}{true}
\bibliographystyle{iopart-num}
\bibliography{main,heavy,smog,LHCb-PAPER,LHCb-CONF,LHCb-DP,LHCb-TDR}

\end{document}